\UseRawInputEncoding
\documentclass[aps,prr,floatfix,twocolumn,longbibliography]{revtex4-2}

\usepackage{amsmath,dcolumn,bm,amsthm,tabularx,subfigure}
\usepackage[colorlinks=true,urlcolor=blue,citecolor=blue,linkcolor=blue]{hyperref}
\usepackage{mathrsfs,times}
\usepackage{bbold,dsfont}
\usepackage{graphicx,graphics,color}
\usepackage{mathtools,nicefrac}
\usepackage{slashed}
\usepackage{amsfonts,amssymb}
\usepackage{braket}
\usepackage{amsmath}
\usepackage{cleveref}
\allowdisplaybreaks

\usepackage[colorlinks=true,
            citecolor=blue,
            linkcolor=blue,
            urlcolor=blue]{hyperref}

\newcommand{\beq}{\begin{equation}}
\newcommand{\eeq}{\end{equation}}
\newcommand{\bea}{\begin{eqnarray}}
\newcommand{\eea}{\end{eqnarray}}

\begin{document}

\title{Local Correlations Determine the Dephasing Threshold of $\mathbb{Z}_2$ Topology}

\author{Rishith Reddy V	}
\email{Rishith Reddy V	}
\affiliation{Department of Physics, Indian Institute of Science Education and Research (IISER) Tirupati, Tirupati 517619, India}

\author{Ashutosh Dubey}
\email{Ashutosh Dubey}
\affiliation{Department of Physics, Indian Institute of Science Education and Research (IISER) Tirupati, Tirupati 517619, India}

\author{Partha Sarathi}
\email{Partha Sarathi}
\affiliation{Department of Physics, Indian Institute of Science Education and Research (IISER) Tirupati, Tirupati 517619, India}

\author{Ankur Das}
\email[Corresponding author.\\]{Ankur Das}
\affiliation{Department of Physics, Indian Institute of Science Education and Research (IISER) Tirupati, Tirupati 517619, India}

\begin{abstract}
In this work, we try to understand the question of how to address the survival of band topology in the presence of dephasing. We investigate this question in the Kane-Mele $\mathbb{Z}_2$ topological insulator. To determine the stability of the topological invariant we construct a single particle  Hamiltonian that also provides identical correlation at the single particle level. The phase transition from topological to trivial is determined by the gap closing of this correlation Hamiltonian. This gap closing is related to purity gap. As a diagnostics we calculated the invariant directly and as a complementary diagnostic, coherent two-terminal transport through the auxiliary correlation Hamiltonian exhibits quantized conductance in the nontrivial region. This does not establish physical conductance under dephasing. Our results identify a predictive local criterion for the survival of single-particle topological structure under a specified noise channel.

\end{abstract}

\maketitle

\noindent \textbf{INTRODUCTION}\\

Topological phases extend the classification of quantum matter beyond spontaneous symmetry breaking~\cite{SciPostPhysLectNotes.11, PhysRevB.107.L041402, PhysRevLett.113.205701, 3g6d-gn7b}, including topological insulators~\cite{RevModPhys.83.1057}, semimetals~\cite{q4gd-b4vb, PhysRevLett.130.186202, SciPostPhysCore.5.1.014, RevModPhys.90.015001, Yan2017, Gao2019}, and superconductors~\cite{bernevig2013topological, Sato2017}. Periodic driving~\cite{Kitamura2022, PhysRevB.107.L121407, PhysRevB.111.035431, PhysRevLett.113.236803, dubey2026topologicalsignaturesquenchdynamics}, strain~\cite{Azizi2025, dccn-czmz, PhysRevB.94.241405}, and other external perturbations~\cite{PhysRevResearch.3.013229, Xu2017, Antonius2016} provide routes to engineering these phases. For gapped noninteracting systems, the classification depends on symmetry and dimensionality~\cite{Ryu2010}, with the Chern number~\cite{PhysRevLett.61.2015} and the $\mathbb{Z}_2$ index~\cite{PhysRevLett.95.146802} characterizing Chern and quantum spin Hall insulators, respectively. Hall response provides a complementary probe that remains useful in interacting systems~\cite{Goldman2024}. Ultracold atoms~\cite{RevModPhys.91.015005, Wang2026, PhysRevLett.107.145301}, photonic systems~\cite{RevModPhys.91.015006, Lan2022, Khanikaev2024}, Floquet synthetic dimensions~\cite{Sridhar2024, PhysRevX.7.041008, vinjamuri2026simulatingmultinodeweylsemimetals}, and superconducting circuits~\cite{Youssefi2022, PhysRevLett.123.080501, PhysRevX.14.041041} provide settings in which to study these phases. Certain topological phases also allow protected qubit encodings~\cite{w2w5-n3lp, Wang2020}, motivating their use in quantum computation~\cite{PhysRevResearch.6.043288}. Topological protection is usually understood in terms of changes to a Hamiltonian: a gapped phase survives symmetry-preserving deformations that do not close the bulk gap. In these settings, however, environmental coupling~\cite{PhysRevLett.89.220405} and measurement backaction~\cite{PhysRevA.64.042110} can reduce coherence and turn pure states into mixed states. The Hamiltonian may retain its band topology throughout this process, while the state's correlations change. The question for an open quantum system~\cite{Rotter2015} is therefore which topological properties of the state survive decoherence, and how their loss can be identified in a mixed-state description~\cite{PhysRevX.8.011035}.

Dissipation can both destabilize topological states and provide a means of preparing them~\cite{PhysRevB.85.174533, coser2019classification, diehl2011topology, Bardyn2013}. A widely studied setting combines quadratic Hamiltonians with Lindblad jump operators that are linear combinations of single-particle creation and annihilation operators~\cite{deng2026dissipation}. For an initially Gaussian state, these dynamics preserve Gaussianity, allowing the evolving state to be fully characterized by its two-point correlations. This provides a tractable framework for investigating pure dark states~\cite{PhysRevA.78.042307}, mixed-state topology, and dissipation-driven topological transitions~\cite{Mao2024}. Dissipative preparation has also been studied for topological phases such as Chern insulators~\cite{PhysRevA.91.042117}. These developments establish that environmental coupling need not simply destroy topology; its effect depends on the structure of the dynamics and the states they produce.

Pure dephasing raises a related question: which topological properties survive when coherence is suppressed without particle loss? Density-based dephasing generally takes an initially Gaussian state outside the Gaussian description, so two-point correlations need not determine the full evolving state. Approaches to this problem include field-theoretical methods~\cite{Sieberer_2016}, R\'enyi correlators, and relative R\'enyi entropies~\cite{pjs3-14cc, nn2m-w4vk}, alongside broader formulations of density-matrix topology~\cite{PhysRevB.91.165140} and diagnostics such as strange correlators~\cite{PhysRevLett.112.247202}. Recent studies have also explored decoherence-induced topology in interacting systems~\cite{chaduteau2026topologydecoherence} and strong-to-weak spontaneous symmetry breaking in dephased Fermi gases~\cite{wang2026observationstrongtoweakspontaneoussymmetry}. Closely related to the present work, Kiely and Xu investigated dephasing-driven changes in the correlation-matrix topology of Chern insulators and three-dimensional time-reversal-invariant topological insulators~\cite{8vtp-khnq}. In this approach, a topological transition requires the closing of the gap around occupation $1/2$ in the bulk correlation spectrum. This gap is similar to the single-particle purity gap~\cite{deng2026dissipation} that one compute commonly for the dissipative quantum system. The resulting invariant characterizes the topology of single-particle correlations, without assuming that these correlations exhaust the information in the mixed many-body state. This distinction is essential when relating bulk correlation topology to boundary and transport diagnostics.

A characteristic feature of topological phases is the presence of boundary states, whose nature depends on the bulk topology, symmetry, and dimensionality. In the integer quantum Hall effect, time-reversal symmetry is broken, and a nonzero Thouless–Kohmoto–Nightingale–den Nijs (TKNN) or Chern invariant~\cite{Fukui2005} is associated with chiral edge states. The quantum spin Hall phase preserves time-reversal symmetry and is characterized by a nontrivial $\mathbb{Z}_2$ invariant~\cite{PhysRevB.75.121403}, with helical edge states comprising counterpropagating modes related by time reversal. Both chiral and helical edge states support charge transport, while the spin structure of helical modes also makes them relevant to spin transport. How decoherence affects the connection between bulk topology and these boundary transport signatures motivates the present study. We investigate a Kane–Mele quantum spin Hall insulator subjected to density dephasing within the quantum master equation. From the dephased single-particle correlation matrix, we compute a bulk $\mathbb{Z}_2$ invariant that distinguishes the trivial and nontrivial correlation phases. Including Rashba spin–orbit coupling allows us to examine this evolution without assuming spin conservation. We then construct the auxiliary correlation Hamiltonian, $H_C=\mathbb{I}/2-C$, and calculate its coherent two-terminal conductance in a finite geometry. The nontrivial correlation topology survives weak dephasing, while stronger dephasing progressively reduces the topological region in the parameter space studied. The region of quantized auxiliary conductance tracks this reduction and eventually vanishes with the nontrivial phase. The agreement between the bulk invariant and the conductance results supports a bulk–boundary connection at the level of the dephased correlation matrix.

\begin{figure*}[t]
\includegraphics[width=1\textwidth]{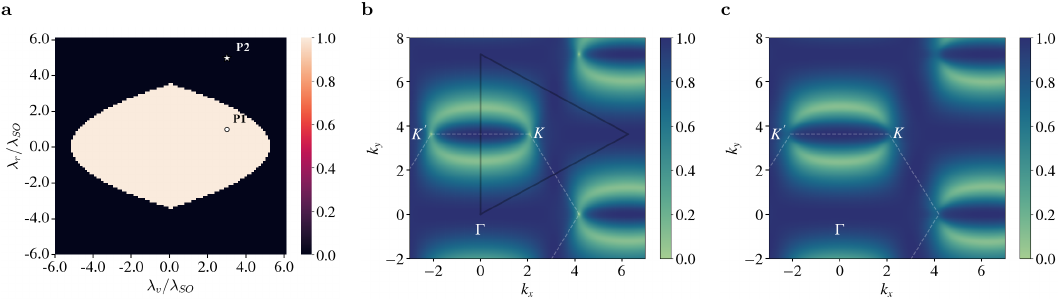}
\caption{\textbf{a}, shows the $Z_2$ phase diagram for the Kane-Mele Model. The central region is the quantum spin Hall phase whereas the outer black region is the trivial phase. \textbf{b} and \textbf{c} show the presence and absence of the Pfaffian zeros in the Brillouin zone for the marked non-trivial (\textbf{P1}) and trivial (\textbf{P2}) phases with $ ({\lambda_v}/{\lambda_{so}},{\lambda_r}/{\lambda_{so}}) = (3,1)$ and $(3,5)$, respectively. The black contour represents the contour over the half BZ around which the $Z_2$ index $I$ is evaluated.
}
\label{Fig:1}
\end{figure*}
\noindent \vbox{} \noindent \textbf{RESULTS}\\

 We consider the Kane–Mele model subjected to density dephasing within the quantum master equation. In this time-reversal-invariant setting, the total Chern number~\cite{Fukui2005} vanishes, while a nontrivial $\mathbb{Z}_2$ index distinguishes the quantum spin Hall phase~\cite{PhysRevB.75.121403}. We first specify the Hamiltonian and dephasing channel, then examine the evolution of the bulk correlation spectrum, its $\mathbb{Z}_2$ topology, and the two-terminal conductance of the associated correlation Hamiltonian.

\noindent \textbf{Kane Mele Model :}\\ 
To begin with, the quantum spin Hall system described by the Kane-Mele model can be viewed as two time-reversed copies of the Haldane model, one for each spin species, with opposite effective magnetic fluxes generated by the complex next-nearest-neighbor hopping terms (to ensure the overall time-reversal symmetry). Furthermore, a Rashba spin–orbit coupling term can be added without breaking the time-reversal symmetry of the system. The corresponding Bloch  Hamiltonian~\cite{PhysRevLett.95.146802} is given by,
\begin{equation}
\begin{aligned}
\mathcal{H}(\bm k) &= \sum_{\alpha = 1}^{5}d_{\alpha}(\bm k)\Gamma^{\alpha} + \sum_{\alpha < \beta = 1}^{5}d_{\alpha\beta}(\bm k)\Gamma^{\alpha\beta},
\end{aligned}
\end{equation}
where the gamma matrices are expressed as a direct product of Pauli matrices accounting for spin and sublattice degrees of freedom. Various gamma matrices are as follows; \(\Gamma^{1}= \sigma^{x}\otimes s^{0}, \Gamma^{2}= \sigma^{z}\otimes s^{0},  \Gamma^{3}= \sigma^{y}\otimes s^{x},  \Gamma^{4}= \sigma^{y}\otimes s^{y},  \Gamma^{5} = \sigma^{y}\otimes s^{z}\). Pauli matrices \(\sigma^{i}\) and \(s^{i}\) account for the sublattice and spin degrees of freedom, respectively. The commutator of gamma matrices is defined as \(\Gamma^{\alpha\beta} = \left[\Gamma^{\alpha}, \Gamma^{\beta}\right]/{2i}\).
The expressions for various coefficients of gamma and their commutator terms involved in the above equation are:
\begin{equation}
\begin{aligned}
d_1(\bm k) &= t(1+2\cos(k_xa/2)\cos(\sqrt{3}k_ya/2)),\\
d_2(\bm k) &= \lambda_\nu,\\
d_3(\bm k) &= \lambda_r(1-\cos(k_xa/2)\cos(\sqrt{3}k_ya/2)),\\
d_{4}(\bm k) &= -\sqrt{3}\lambda_{r}\sin(k_{x}a/2)\sin(\sqrt{3}k_{y}a/2),\\
d_{12}(\bm k) &= -2t\cos(k_{x}a/2)\sin(\sqrt{3}k_{y}a/2),\\
d_{15}(\bm k) &= \lambda_{so}(2\sin(2k_{x}a/2) - 4\sin(k_{x}a/2)\cos(\sqrt{3}k_{y}a/2)),\\
d_{23}(\bm k) &= -\lambda_{r}(\cos(k_{x}a/2)\sin(\sqrt{3}k_{y}a/2)),\\
d_{24}(\bm k) &= \sqrt{3}\lambda_{r}(\sin(k_{x}a/2)\cos(\sqrt{3}k_{y}a/2)).
\end{aligned} 
\end{equation}
Various gamma matrices are even under the time reversal symmetry i.e. \(\mathcal{T}\Gamma^{a}\mathcal{T} = \Gamma^{a}\), while all the commutator terms of the gamma matrices are odd under the time reversal symmetry, i.e \(\mathcal{T}\Gamma^{ab}\mathcal{T} = -\Gamma^{ab}\). The time reversal symmetry operator is given by: \(\mathcal{T} = i\left(\mathbb{I} \otimes s^{y}\right)\mathcal{K}\), where \(\mathcal{K}\) is the conjugation operator. Also, the coefficients of gamma matrices are symmetric under the transformation \(\bm k \to -\bm k\) i.e., \(d_{a}(\bm k) = d_{a}(-\bm k) \), whereas the coefficients of commutator terms are antisymmetric under the time-reversal symmetry, i.e., \(d_{ab}(\bm k) = -d_{ab}(-\bm k) \). As a result of this Hamiltonian of Kene Mele model respects the time reversal symmetry.\\

In the absence of the Rashba spin-orbit coupling term \(\lambda_{r} = 0\), the \(S_z\) component of spin is conserved. Although the Haldane model for each spin species individually breaks time-reversal symmetry, the full Hamiltonian remains time-reversal symmetric. Since there is no mixing between the two spin sectors, each spin sector can be treated independently. Consequently, one can compute the TKNN invariant, or equivalently the spin-resolved Chern numbers, \((c_{\uparrow}, c_{\downarrow})\), for the Haldane model corresponding to each spin species. Because the full Hamiltonian preserves time-reversal symmetry, the total Chern number vanishes, \(c = c_{\uparrow} + c_{\downarrow} = 0\). However, the difference between the Chern numbers of the two spin sectors defines a quantized topological invariant, which is directly related to the quantized spin Hall conductance.\\

In the presence of Rashba spin-orbit coupling \(\lambda_{r} \ne 0\), the \(S_{z}\) component is no longer conserved. This term mixes the spin-up and spin-down components. However, it still preserves the time-reversal symmetry of the Hamiltonian. In the presence of this term, the spin sectors can no longer be treated independently, and the spin-resolved Chern numbers \((c_{\uparrow}, c_{\downarrow})\) are no longer well-defined. Consequently, a different topological invariant often called as $\mathbb{Z}_2$ index is required to characterize the topological phase of the system. We discuss this invariant in the following section in more detail. \\

\begin{figure*}
\includegraphics[width=\textwidth]{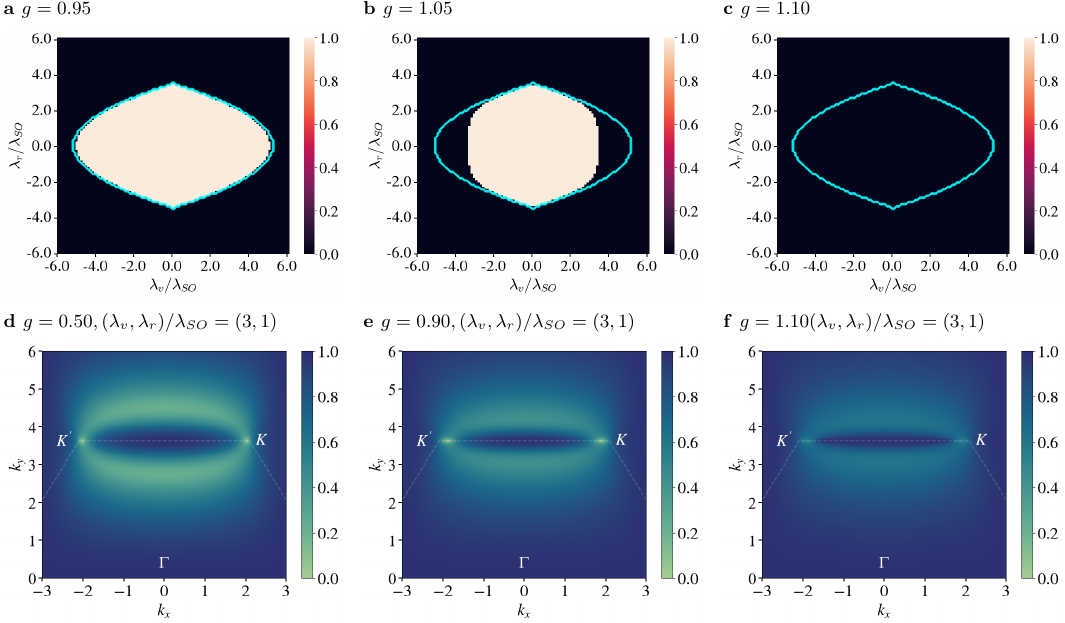}
\caption{\textbf{Evolution of Topological Phases}. \textbf{a} - \textbf{c} show that the pure state boundary (cyan contour) is robust until dephasing rate $g=0.95$, after which the states near the boundary transition, eventually losing all topology near $g=1.1$. \textbf{d} - \textbf{f} show the evolution of the Pfaffian zeros in the chosen topological case, eventually vanishing at strong dephasing $g = 1.1$. The trivial case, as expected, shows no such transition. }
\label{fig:pfaffian-dep}
\end{figure*}

\noindent \textbf{Clean Limit:}\\
The above Bloch Hamiltonian describes a four-band model. 
When the system is free from decoherence noise, in the absence of Rashba spin-orbit coupling, the gap is given by $\Delta = |6\sqrt{3}\lambda_{SO} - 2\lambda_{v}|$. In the limit, $\lambda_{v} \ge 2\sqrt{3}\lambda_{SO}$, system behaves as an insulator and gap is dominated by $\lambda_{v}$, whereas in the limit $3\sqrt{3}\lambda_{SO} \ge \lambda_{v}$, system exhibits quantum spin Hall phase. In the presence of Rashba spin orbit coupling term, $S_{z}$ component is no longer conserved, and region exhibiting the spin Hall phase shrinks as shown in Fig.~\ref{Fig:1}(a). Even in the presence of the Rashba term, for the limit \(\lambda_{r} \le \sqrt{3}\lambda_{so}\), the region of the phase diagram is topologically nontrivial and is adiabatically connected to the quantum spin Hall phase for \(\lambda_{r} = 0\).
Further, the non-trivial quantum spin Hall phase is robust and persists even if $C_{3}$ symmetry is removed.

Time-reversal symmetry classifies the points of the Brillouin zone into \textit{even and odd subspaces}. This division depends on how the occupied subspace of the Bloch Hamiltonian, $\{\ket{u(\bm{k})}\}$, is related to its corresponding time-reversal-symmetric partner. In the case of the even subspace, the original occupied states and their time-reversal partners are related by a $U(2)$ transformation. Thus, the time-reversed partner states belong to the same original occupied subspace. The center of the Brillouin zone, as well as the centers of the edges of the Brillouin zone, satisfy the condition for the even subspace. In the case of the odd subspace, the original occupied subspace and its time-reversed partner subspace are orthogonal to each other, and thus the time-reversed states do not belong to the original occupied subspace.

The topology of the Kane-Mele model is established using the $Z_2$ classification~\cite{PhysRevLett.95.146802}, which in turn involves identifying these subspaces.  The $Z_2$ index can be determined by counting the number of pairs of complex zeros of the complex function $P(\bm{k})$. In the case of the Kane-Mele model, this complex function is the Pfaffian of the $2\times2$ matrix associated with the time-reversal operator in the basis of the occupied subspace. For a $2\times2$ matrix, the Pfaffian is the off-diagonal element, so we can write,
\begin{equation}
 P(\bm{k}) = \bra{u_{1}(\bm k)} \mathcal{T} \ket{u_{2}(\bm k)}, 
 \label{eq:pf}
\end{equation}
where \((\ket{u_{1}(\bm k) }, \ket{u_{2}(\bm k)})\) is occupied single particle eigen state.
In the honeycomb lattice, zeros of the Pfaffian occur at high-symmetry points $\bm K$ and $-\bm K$ of the honeycomb lattice Brillouin zone as shown in Fig.~\ref{Fig:1}(b), whereas in the case of the trivial phase, zeros of the Pfaffian disappear from the high-symmetry points as shown in Fig.~\ref{Fig:1}(c).
In the topological phase, the $\mathbb{Z}_2$ topological index can be calculated by evaluating the winding of the phase of \(P(\mathbf{k})\) along a loop (shown by the black line in Fig.~\ref{Fig:1}(b)) in half of the Brillouin zone. The half Brillouin zone is defined such that \(\mathbf{k}\) and \(-\mathbf{k}\) are never included simultaneously. The loop encloses one zero of the Pfaffian, and the $\mathbb{Z}_2$ index is obtained by taking the winding number modulo 2:
\begin{equation}
I = \frac{1}{2\pi i}\oint_C d \bm k\cdot\nabla_{\bm k}
\log\left[P(\bm k)+i\delta\right].
\label{eq:z2}
\end{equation}\\
In the above equation, \(\delta\) is the convergence factor. More particularly, the integral I mod 2 gives the $\mathbb{Z}_2$ topological invariant. In the case of quantum spin hall phase, $\mathbb{Z}_2$ index is quantize to 1, where as for the trivial phase, $\mathbb{Z}_2$ is 0. The phase diagram is shown in Fig.~\ref{Fig:1}(a). The transition from the quantum spin hall phase to the trivial phase has sharp boundaries where band closing and reopening occur in the spectrum of the Hamiltonian.  \\

\begin{figure*}[t]
\centering
\includegraphics[width=\linewidth]{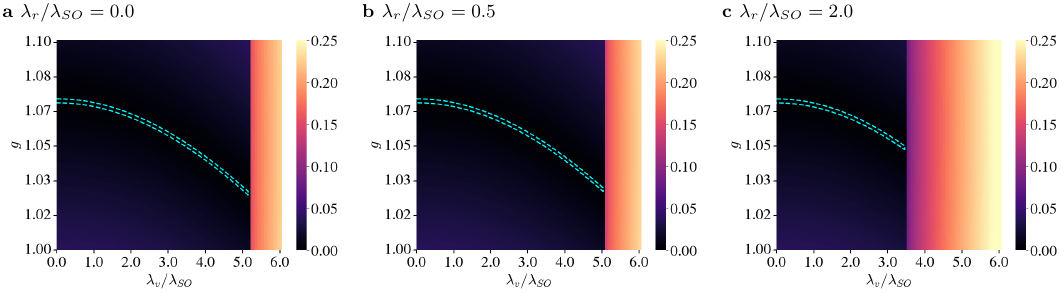}
\caption{\textbf{Gap closure of the flat band Hamiltonian}. The plots show the closure of the effective flat band Hamiltonian gap corresponding to the exact transition points $(\lambda_v /\lambda_{SO}, \lambda_r/\lambda_{SO}, g)$ from the topological to trivial phase. }
\label{purities}
\end{figure*}

\noindent \textbf{De-phased  system:}\\
In the presence of decoherence noise, the phase diagram changes drastically depending upon the strength of noise. To investigate this effect, we construct the single-particle correlation matrix: $   C_{ij}^{\alpha\beta} = \text{Tr}(\rho \, c_{i,\alpha}^\dagger c_{j,\beta})$, where ${i,j}$ are the spatial indices, ${\alpha, \beta}$ denote the internal degree of freedom like spin, sublattice, etc. t.c and $\rho$ is the density matrix of the clean system. In the absence of decoherence noise, the density matrix of the system is pure, i.e $\rho = \ket{\psi}\bra{\psi}$, where $\ket{\psi}$ is the many-body ground state. Owing to the non-interacting nature of the system, the many-body ground state is obtained
by filling all the single-particle states below the Fermi energy.
With this, we find that the correlation matrix is a projector on the
occupied single-particle states. From the correlation matrix, we construct an effective (flat) Hamiltonian $H_{C} = \mathbb{I} - \frac{C}{2}$. 
Working with the flat Hamiltonian is reasonable because it is topologically equivalent to the original Bloch Hamiltonian.
We refer to the Methodology section for the details of the calculation.
We evolve the flat Hamiltonian as a function of the decoherence strength. To accomplish this, we compute the single-particle correlation matrix as a function of decoherence strength using Eq~\eqref{eq:dep-corr}. By diagonalizing the flat Hamiltonian, we obtain a pair of dephased occupied single-particle eigenstates. Together with the time-reversal operator, these dephased occupied states allow us to compute the Pfaffian in the presence of decoherence noise using Eq.~\eqref{eq:pf}.  We find that, for a point in parameter space where the zeros of the Pfaffian are located at the corners of the Brillouin zone, as shown in Fig.~\ref{Fig:1}(b), these zeros disappear at sufficiently strong strength of decoherence noise, as shown in Fig.~\ref{fig:pfaffian-dep}(f). Importantly, the disappearance of the Pfaffian zeros is not gradual. Instead, the zeros remain robust against weak decoherence, as shown in Fig.~\ref{fig:pfaffian-dep}(d,e). The disappearance of the Pfaffian zeros signals a change in the topological phase diagram and the associated transition induced by decoherence. 

To verify the effect of decoherence noise on the phase diagram, we compute the $\mathbb{Z}_2$ topological index using Eq.~\eqref{eq:z2} and the Pfaffian made out of dephased occupied states, and construct the corresponding phase diagram as a function of \(\lambda_{v}/\lambda_{so}\), \(\lambda_{r}/\lambda_{so}\), for various values of the decoherence strength \(g\). Both for clean as well as dephased system we use Z2pack~\cite{gresch2017z2pack, PhysRevB.83.235401}, for computing the integral given by Eq.~\eqref{eq:z2} and thus for the computation of $\mathbb{Z}_2$ topological invariant. We explicitly show the phase diagram for \(g = 0.95, 1.05\) and \(g = 1.1\) in Fig.~\ref{fig:pfaffian-dep}(a,b,c) respectively. We find that the topological region in the phase diagram remains unchanged for decoherence strength \(g = 0.95\). Beyond this value, the topological region shrinks, predominantly along the \(\lambda_{v}\) direction, as observed for \(g=1.05\). Upon further increasing the decoherence strength to \(g=1.1\), the topological region completely disappears. These results demonstrate that sufficiently strong decoherence can destroy the topological phase. The shrinking of the topological region in the phase diagram indicates that the effect of decoherence noise becomes significant beyond a certain strength. However, gradual shrinking of the phase diagram takes place in a very small window of decoherence noise strength, i.e approximately \(g\in(1.0,1.1)\), driving the system from a topologically non-trivial phase to a topologically trivial phase.

From the conventional understanding of topological phase transitions in the closed system, such a transition is generally accompanied by the closing and reopening of a spectral gap of the Hamiltonian. In the case of an open quantum system, one can study such a transition by keeping track of the opening and closing of the purity gap (spectral gap of the correlation matrix). In this case we investigate this, by computing the correlation-matrix gap and find that, for a fixed value of \(\lambda_{r}\) as shown in Fig.~\ref{purities}. The gap closes within the range \(g\in(1.0,1.1)\), consistent with the topological phase diagram discussed above. Correlation matrix gap closing also helps in estimating the critical value of decoherence noise strength $g$, at which the system undergoes a transition from the nontrivial (QSH) phase to the trivial phase at a given value of \((\lambda_{v}/\lambda_{so}, \lambda_{r}/\lambda_{so})\)
. We observe that for larger values of \(\lambda_{v}\), a comparatively weaker decoherence strength is sufficient to drive the topological phase transition. Furthermore, with increasing value of \(\lambda_{r}\) , the minimum value of decoherence noise strength required for topological phase transition at larger value of \(\lambda_{v}\) increases. However, for the minimum value of \(\lambda_{v}\), the magnitude of \(g\), required is almost independent of the magnitude of \(\lambda_{r}\). This indicates that the topological phase becomes increasingly susceptible to decoherence as \(\lambda_{v}\) is increased.
This also explain the shrinking of phase diagram more in \(\lambda_{v}\) direction in comparison to \(\lambda_{r}\) direction as shown in Fig.\ref{fig:pfaffian-dep}(b).\\ 

\noindent \textbf {Two-terminal transport on dephased flat Hamiltonian :}\\

\begin{figure}[t]
\includegraphics[width=\columnwidth]{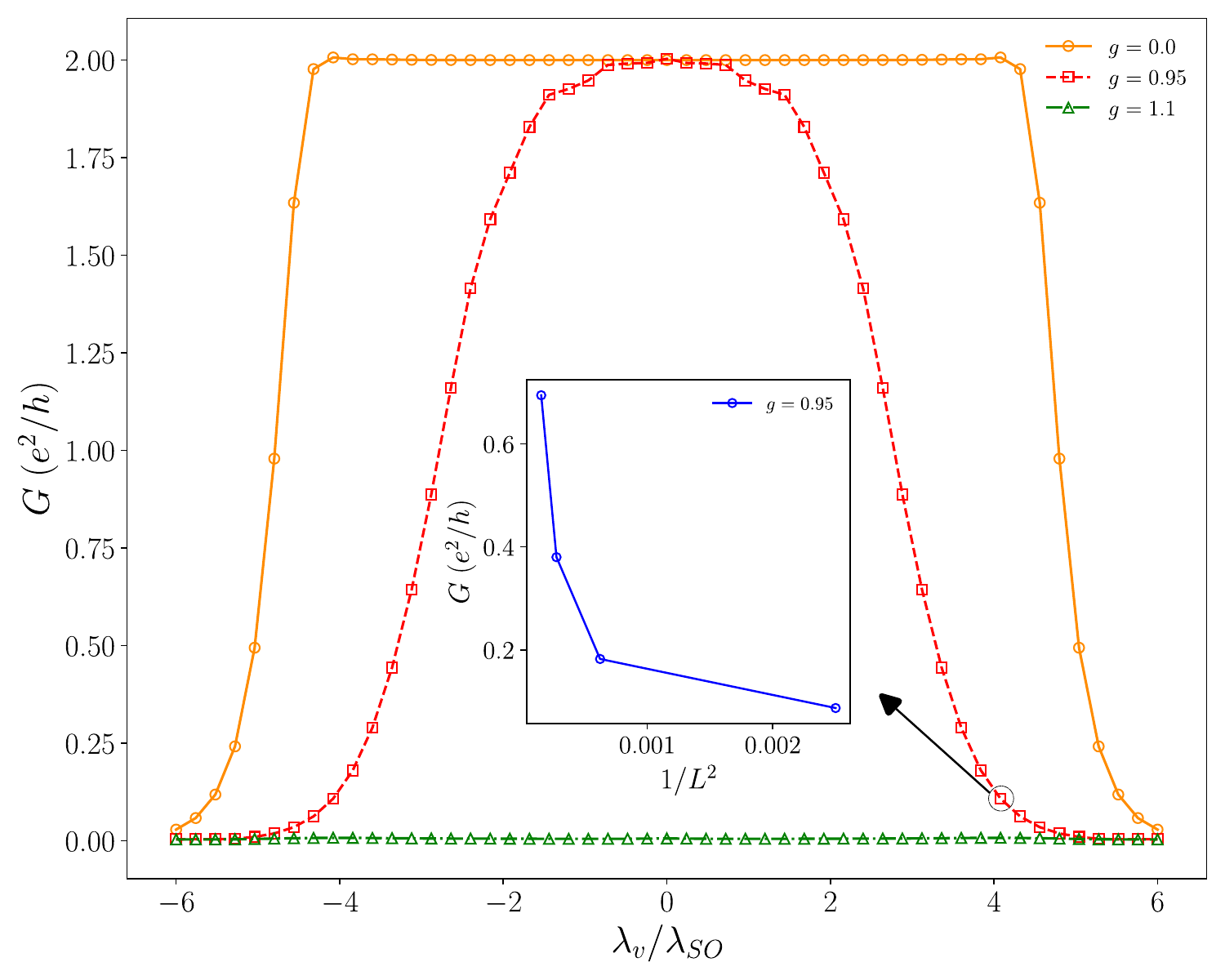}
\caption{\textbf{Two terminal conductance}:Two-terminal conductance($G(e^2/h)$) vs $\lambda_v$ for three different values of $g$ for a scattering region of size $30 \times 30$.The figure is in agreement with the Pfaffian phase diagram shown in \cref{fig:pfaffian-dep} which shows a shrinking topological phase.System size scaling of $G$ at $\lambda_r=0.5\lambda_{so}$ and $\lambda_v=4\lambda_{so}$(inset)shows that the non-quantized values of the two terminal conductance is a consequence of finite-size effect and we claim that at higher system size the conductance becomes quantized.}
\label{fig:2_term_conductance}
\end{figure}

Unlike a Chern insulator, where time-reversal symmetry is broken and the topological phase is characterized by the TKNN invariant (Chern number), a quantum spin Hall (QSH) insulator, such as the one described by the Kane–Mele model, preserves time-reversal symmetry and is generally characterized by a $\mathbb{Z}_2$ invariant. The Chern number can be extracted from the quantized bulk Hall conductivity, but the $\mathbb{Z}_2$ invariant cannot be directly determined from the charge Hall conductivity.

In the absence of the Rashba spin-orbit coupling, the Kane–Mele model has spin-rotation symmetry. In this case, the spin-up and spin-down sectors can be treated separately, and the topological and trivial phases can be distinguished using the spin-resolved Chern number. The spin-resolved Chern number can, in turn, be related to the quantized spin-resolved Hall conductivity. However, in the presence of Rashba spin-orbit coupling, the spin-rotation symmetry is broken. As a result, the $\mathbb{Z}_2$ topological phase cannot be identified from a quantized bulk spin Hall conductivity. This can be understood by considering the spin accumulation at the edge of the sample. The quantum spin Hall conductance is related to the rate of spin accumulation at the edge, which can be obtained from the expectation value of \(dS_z/dt\) for the left- and right-moving single-particle eigenstates that cross the bulk gap and form the helical edge states. In a realistic system, Rashba spin-orbit coupling is generally present. In this case, the left- and right-moving edge states cannot be assigned a definite spin quantum number because they are no longer eigenstates of \(S_z\). Consequently, the spin Hall conductance is no longer quantized. It is important to note, however, that although the spin Hall conductance is not quantized in the presence of Rashba spin-orbit coupling, it remains nonzero in the QSH phase, while it vanishes in the trivial insulating phase.

Usually, in a QSH system, a quantized two-terminal transport measurement provides a good estimate of the topological phase, even in the presence of Rashba spin-orbit coupling. In this work, we use two-terminal transport to estimate the topological phase of a dephased QSH system. We begin with the dephased flat Hamiltonian and Fourier transform it from momentum space to real space with open boundary conditions. We then attach metallic leads to the two opposite sides of the system and calculate the conductance in a two-terminal setup using Eq.~\eqref{eq:two-cond}. The details of the calculation and the setup are given in the Methodology section.

For a fixed Rashba spin-orbit coupling, $\lambda_r = \lambda_{so}/2$, and decoherence noise strength, $g = 0.95$, we find that the two-terminal conductance is quantized over the range of $\lambda_v$ for which the $Z_2$ invariant is nontrivial, as shown in Fig.~\ref{fig:2_term_conductance}(a). In this range, the conductance remains quantized in the clean system as well. In principle, the conductance at $g = 0.95$ should match its clean-limit value, as indicated by Fig.~\ref{fig:pfaffian-dep}(a). However, at larger values of $\lambda_v$, we observe a deviation from the clean-limit conductance. This deviation arises from finite-size effects. Despite these effects, our two-terminal transport results indicate that the topological phase remains robust against decoherence at $g = 0.95$.

In contrast, at a higher decoherence noise strength of $g = 1.1$, the two-terminal conductance becomes zero over the range of $\lambda_v$ for which the clean system is topological. This result demonstrates that decoherence can drive a topological phase transition.

To further investigate finite-size effects, we consider a point near the boundary of the topological phase diagram, specified by $\lambda_v = 4\lambda_{so}$ and $\lambda_r = 0.5\lambda_{so}$. At this point, the clean system exhibits a quantized two-terminal conductance of $2e^2/h$. However, in the presence of decoherence with $g = 0.95$, the conductance deviates from this quantized value due to finite-size effects. We find that the conductance approaches its quantized value as the system size increases, indicating that it tends towards quantization in the thermodynamic limit. Overall, the system-size dependence of the two-terminal conductance shows that transport measurements in the presence of decoherence are consistent with the topological phase predicted by the $Z_2$ invariant of the dephased flat-band Hamiltonian.

\vbox{} \noindent \textbf{DISCUSSION}

\noindent In this work, we have investigated how density dephasing affects the connection between bulk $\mathbb{Z}_2$ correlation topology and boundary transport signatures in the Kane--Mele quantum spin Hall insulator. The dephased single-particle correlation matrix retains a nontrivial index over a finite interval of dephasing strength, before undergoing a transition associated with the closing of its single-particle purity gap. We further find that the region of quantized two-terminal conductance of the auxiliary correlation Hamiltonian tracks the nontrivial bulk region as dephasing increases. This correspondence persists in the presence of Rashba spin--orbit coupling over the parameter ranges studied, including cases where spin is not conserved. The finite-dephasing stability is consistent with the results of Kiely and Xu~\cite{8vtp-khnq}; our study adds an explicit comparison between the bulk correlation invariant and a boundary transport diagnostic in a two-dimensional quantum spin Hall system. In their clean Chern-insulator example, the critical dephasing strength diverges logarithmically near a particular band-inversion point. A notable feature of our results is that the critical strengths cluster within the narrow interval $g_c\in(1.02,1.08)$ across the parameter sets investigated, as illustrated in Fig.~\ref{purities}. Although $\lambda_v$ and $\lambda_r$ affect the extent of the topological phase, the dephasing strength required to eliminate its correlation topology remains comparatively \emph{insensitive}~ to these variations. This clustering suggests a nearly universal dephasing scale within the regime studied, providing a quantitative measure of the stability of the $\mathbb{Z}_2$ correlation topology under the chosen noise channel. Understanding the origin and range of validity of this weak parameter dependence is an important question raised by our results.

The transition studied here concerns the topology of the evolving state’s single-particle correlations. The initial Kane--Mele Hamiltonian specifies the prepared state, while the subsequent evolution is governed solely by density dephasing. The relevant gap closing therefore occurs in the correlation spectrum, without requiring a change in the band topology of the preparation Hamiltonian. Within this description, the phase diagrams reveal a distinction between the extent of the nontrivial region and the dephasing strength at which it disappears. At fixed $g$, the phase boundary in $\lambda_v$ depends on $\lambda_r$, and increasing dephasing progressively removes the nontrivial region, particularly at larger $\lambda_v$. Our numerical sweeps also indicate that increasing $\lambda_r$ narrows the interval of critical dephasing strengths. Thus, the clustering of $g_c$ coexists with a phase boundary that remains sensitive to the model parameters. The nearly common dephasing scale is consequently a feature of the sampled transitions, while the full phase diagram retains information about the underlying Hamiltonian.

A central question raised by our results is whether the narrow spread of critical dephasing strengths reflects a more general constraint on the stability of correlation topology. Identifying which features of the initial state and noise channel determine this scale would provide a predictive criterion for noise tolerance and clarify the range of validity of the nearly universal behavior observed here. Spin-selective density dephasing offers a way to investigate how the symmetry of the noise affects the accessible phases. In particular, unequal dephasing rates for spin-up and spin-down particles break the time-reversal symmetry of the channel, raising the possibility of generating a nonzero Chern invariant of the correlation matrix. Establishing whether this occurs would clarify whether selective decoherence can serve as a means of controlling correlation topology. Combining dephasing with periodic driving, such as circularly polarized light, would address the stability of topology generated dynamically in an initially trivial system. Finally, incorporating coherent Hamiltonian evolution and disorder would test whether the stability found under the present protocol persists when dephasing competes with the dynamics and spatial inhomogeneity of the system.

Cold-atom quantum simulators offer an experimentally accessible route to investigating these transitions through controlled noise and measurements of single-particle correlations. Density dephasing could be introduced by applying stochastic onsite potentials and averaging over repeated noise realizations, with the dephasing strength varied through the noise amplitude or exposure time. To reproduce the channel studied here, the fluctuations should couple identically to the spin and sublattice degrees of freedom within each unit cell and independently between cells, with coherent evolution suppressed during the dephasing step. A useful experimental feature is that locating the transition requires only single-particle correlations, even when the evolving many-body state is non-Gaussian. Existing time-of-flight schemes for probing topological textures and currents provide a starting point~\cite{alba2011seeing, rubio2020seeing}. Combined with suitable control and readout of the spin and sublattice degrees of freedom, momentum-resolved measurements could reconstruct the correlation spectrum and reveal its gap closing at occupation $1/2$. Such measurements would directly test the predicted stability interval and the weak parameter dependence of the critical dephasing strength.

\vbox{} \noindent \textbf{METHODS}\\
\noindent \textbf{Effective Flat Hamiltonian :}\\
The evolution of the density matrix of the quantum system subject to dephasing noise is modeled using a Lindblad master equation of the form :
\begin{equation}
    \partial_g \rho = \sum_l \left( L_l \rho L_l^\dagger - \frac{1}{2} \{\rho, L_l^\dagger L_l\} \right)
\label{eq:methods-1}
\end{equation}
where $l$ is the site index and the sum is over all the site. Further we are interested in the effect of dephasing noise, we ignore the Hamiltonian term in the master equation that accounts for the coherence dynamics. We choose the jump operator to account for the case of density dephasing $L_i =  n_i$. In other perspective we can think of ancilla degree of freedom couples with the density operator at a given site. And further we assume jump operator is independent of internal degree of freedom like spin, sublattice, e.t.c. We compute the correlation matrix given by $   C_{ij}^{\alpha\beta} = \text{Tr}(\rho \, c_{i,\alpha}^\dagger c_{j,\beta})$, where ${i,j}$ are the spatial indices, and \({\alpha, \beta}\) are indices for internal degree of freedom. In the expression of correlation matrix, trace is over many body Hilbert space and density matrix is given by \(\rho = \ket{\psi}\bra{\psi}\), where where \(\ket{\psi}\) is the many body ground state. Initially, in the absence of decohorence noise, the density matrix of the system is pure. Further the correlation matrix can be written as : \(\mathcal{C}_{ij} = \bra{\psi} c_{i}^{\dagger}c_{j}\ket{\psi}\). Owing to the fact that system is non interacting, the many body ground state is obtained by filling all the single particle state below the Fermi energy. With this we find that correlation matrix is projector on the occupied single-particle states. In this work study the evolution of correlation matrix as a function of decohorence noise. The evolution
of the correlation function is given by the adjoint master equation, which for
an arbitrary operator \(\hat{O}\), reads (only in the case of time independent Lindblad evolution, the time dependent case requires more information) as :
\begin{align}
\partial_g \hat{O}
= \sum_i
\left(\hat{n}_i \hat{O} \hat{n}_i -  \frac{1}{2}\left\{\hat{n}_i,\hat{O}\right\}\right).
\end{align}
Similar to the case of evolution of density matrix given in Eq.~\eqref{eq:methods-1},  we ignore the Hamiltonian term from the adjoint master equation as we are only interested in the decoherence dynamics. Thus similar equation for the correlation matrix is given by :
\begin{align} 
\partial_g C_{ij}  &= Tr\left(\rho \left( \sum_l \hat{n}_l \hat{c}_i^\dagger \hat{c}_j \hat{n}_l - \frac{1}{2}\{\hat{n}_l, \hat{c}_i^\dagger \hat{c}_j\} \right)\right) .
\end{align}
On expressing the density operator $\hat{n}_{i}$ in terms of single particle creation and annhilation operator, the first order differential equation describing the evolution of element of correlation matrix as a function of decohorence noise strength is given by,
\begin{align} 
\notag  \partial_g C_{ij}  = \sum_l \Bigg( &\delta_{il} \text{Tr}(\rho c_l^\dagger c_j c_l^\dagger c_l) - \delta_{lj} Tr(\rho c_i^\dagger c_l c_l^\dagger c_l) \\ + \text{Tr}(\rho c_i^\dagger c_j c_l^\dagger c_l) 
&- \frac{1}{2} Tr(\rho c_l^\dagger c_l c_i^\dagger c_j) - \frac{1}{2} Tr(\rho c_i^\dagger c_j c_l^\dagger c_l) \Bigg) .
\label{eq:methods-7}
\end{align}
In writing the above equation we use the identity,
\begin{align}
\hat{n}_l \hat{c}_i^\dagger \hat{c}_j &= \hat{c}_l^\dagger \hat{c}_l \hat{c}_i^\dagger \hat{c}_j = \delta_{il} \hat{c}_l^\dagger \hat{c}_j - \delta_{lj} \hat{c}_i^\dagger \hat{c}_l + \hat{c}_i^\dagger \hat{c}_j \hat{c}_l^\dagger \hat{c}_l. 
\end{align}
The last three terms in Eq.~\eqref{eq:methods-7} can be rewritten simply as (upon using the identity on the middle term),
\begin{equation}
   \text{Tr}(\rho c_i^\dagger c_j c_l^\dagger c_l) \left(1 - \frac{1}{2} - \frac{1}{2}\right) - \frac{1}{2} \delta_{il} Tr(\rho c_l^\dagger c_j) + \frac{1}{2} \delta_{lj} Tr(\rho c_i^\dagger c_l) \nonumber 
\end{equation}
\begin{equation}
       = -\frac{1}{2} \delta_{il} C_{lj} + \frac{1}{2} \delta_{lj} C_{il}. 
\end{equation}
which upon summing over $l$ goes to zero. The remaining terms are then given by,
\begin{equation}
\partial_g C_{ij} = \sum_l \left( \delta_{il}\delta_{lj} Tr(\rho c_l^\dagger c_l) - \delta_{lj} Tr(\rho c_i^\dagger c_l) \right).
\end{equation}
which upon summing over $l$ gives us,
\begin{align} 
\partial_g C_{ij} = (\delta_{ij} - 1)C_{ij} .
\end{align}
As in our case, dephasing noise doesn't break the translational symmetry of the quantum system, momentum \(\bm k\) is good quantum number. After Fourier transform we get,
\begin{equation}
\partial_g C^{\alpha\beta}(\boldsymbol{k}) = -C^{\alpha\beta}(\boldsymbol{k}) + \int _{BZ}\frac{d^2 k'}{(2\pi)^2} C^{\alpha\beta}(\boldsymbol{k}').
\end{equation}
Above equation is the first order linear differential equation of the momentum dependent correlation matrix. The evolution of the correlation matrix as a function of $g$ is then explicitly given by : 
\begin{equation}
    C^{\alpha\beta}(\boldsymbol{k}, g) = e ^{-g} C^{\alpha\beta}(\boldsymbol{k}, 0) + (1 - e^{-g}) \Lambda
\label{eq:dep-corr}
\end{equation}
where $\Lambda = \int_{BZ} \frac{d^2 k'}{(2\pi)^2} C^{\alpha\beta}(\boldsymbol{k}',0)$. One should note that the integral term in the above differential equation is the constant of motion which can be seen as,
\begin{equation}
\begin{aligned}
 \frac{1}{(2\pi)^2}\int d^2k\,\partial_g C(k) &=
-\frac{1}{(2\pi)^2}\int d^2k\,C(k)\\
&+
\frac{1}{(2\pi)^2}\int d^2k
\int\frac{d^2k'}{(2\pi)^2}C(k').
\end{aligned}
\end{equation}
which implies
\begin{equation}
\partial_g
\left(
\int\frac{d^2k}{(2\pi)^2}C(k)
\right)
=
-\int\frac{d^2k}{(2\pi)^2}C(k)
+
\int\frac{d^2k}{(2\pi)^2}C(k)
=0
\tag{18}
\end{equation}

Therefore, the integral term is a constant and can be evaluated using any $C(k,g)$; we choose it to be specified as $C(k,0)$, the projector onto the occupied ground state of any specified Hamiltonian (needs to be translationally invariant). From the correlation matrix, for any value of noise strength, one can construct the effective (flat) Hamiltonian for further study as:
\begin{equation}
 \mathcal{H}_{C}(\bm k, g) = \mathbb{I} - \frac{C(\bm k,g)}{2}. \label{eq:flat} 
\end{equation}

The momentum dependent correlation matrix can be used as a topological marker to observe the phase diagram and the phase transition in the presence of decoherence n noise. One should note the integral term appear in the above equation is actually constant of motion. This can be easily understood by noting that \\

\noindent \textbf{Two-terminal transport through the correlation Hamiltonian}\\

We calculate the coherent two-terminal conductance of the correlation Hamiltonian to examine the transport signatures associated with its bulk $\mathbb{Z}_2$ topology. At each dephasing strength $g$, the dephased correlation matrix defines a static auxiliary Hamiltonian, $\mathcal{H}_{C}(\bm{k},g)$, given in Eq.~\eqref{eq:flat}. We construct a finite scattering region from this Hamiltonian and connect it to two semi-infinite metallic leads. The conductance is then obtained using the Landauer--B\"uttiker formalism~\cite{Datta_1995}.

We first transform the bulk correlation Hamiltonian to real space. In a periodic Bloch basis defined with respect to the unit-cell coordinates,
\begin{equation}
    \mathcal{H}_{\bm R,\bm R'}(g)
    =
    \frac{1}{N_k}\sum_{\bm k}
    e^{i\bm k\cdot(\bm R-\bm R')}
    \mathcal{H}_{C}(\bm k,g),
\end{equation}
where the sum runs over a uniform mesh of $N_k$ points in the Brillouin zone. The lattice vectors are
$\bm R=n_1\bm a_1+n_2\bm a_2$, where $n_1,n_2\in\mathbb Z$ and $\bm a_1,\bm a_2$ are the primitive translation vectors. Each matrix element $\mathcal{H}_{\bm R,\bm R'}$ acts on the sublattice and spin degrees of freedom within the unit cell.

\begin{figure}[h]
\centering
\includegraphics[width=\linewidth]{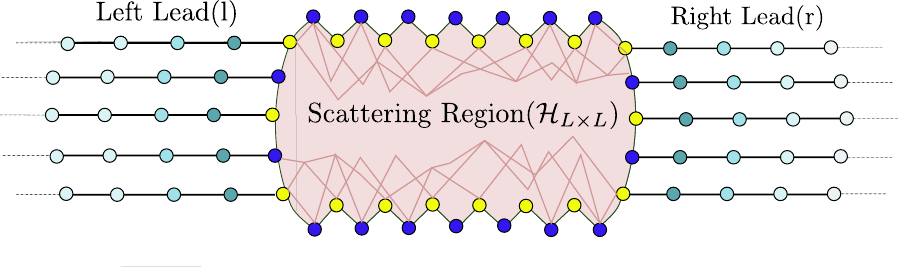}
\caption{\textbf{Two-terminal scattering geometry.}
Blue and yellow circles denote the two sublattices of the honeycomb lattice. The central region is described by the finite correlation Hamiltonian $\mathcal{H}_{L\times L}$ and is connected to semi-infinite conducting leads on the left and right. Lines within the scattering region illustrate the longer-range hopping terms generated by Fourier transforming $\mathcal{H}_{C}(\bm k,g)$.}
\label{Fig:2_terminal_setup}
\end{figure}

The finite scattering region is obtained by restricting both $\bm R$ and $\bm R'$ to a domain $\Omega_L$ containing $L\times L$ unit cells:
\begin{equation}
    \mathcal{H}_{L\times L}(g)
    =
    \sum_{\bm R,\bm R'\in\Omega_L}
    \psi_{\bm R}^{\dagger}
    \mathcal{H}_{\bm R,\bm R'}(g)
    \psi_{\bm R'}.
\end{equation}
Here $\psi_{\bm R}$ is the four-component annihilation operator in the sublattice--spin basis. Restricting the unit-cell coordinates defines the finite geometry with open boundaries. The resulting matrix elements are assigned to the corresponding sites of the honeycomb lattice. The Fourier-transformed correlation Hamiltonian can contain hopping terms beyond those present in the original Kane--Mele Hamiltonian. The scattering region and its connections to the leads are illustrated in \cref{Fig:2_terminal_setup}.

At a fixed energy $E$, a scattering state associated with an incoming propagating channel $n$ can be expanded within the uniform leads as
\begin{equation}
    \Phi_n(i)
    =
    \psi_n^{I}(i)
    +
    \sum_m S_{mn}(E)\psi_m^{O}(i)
    +
    \sum_p \widetilde S_{pn}(E)\psi_p^{E}(i).
\end{equation}
The functions $\psi^{I}$ and $\psi^{O}$ denote incoming and outgoing propagating lead modes, normalized to carry unit current, while $\psi^{E}$ denotes evanescent modes that decay away from the scattering region. The index $i$ labels the lattice and internal degrees of freedom. Evanescent modes contribute to matching the wave function at the contacts but carry no asymptotic transmitted current. The scattering matrix $S$ relates the amplitudes of the incoming and outgoing propagating modes.

For the two-lead geometry,
\begin{equation}
    S(E;g)
    =
    \begin{pmatrix}
        r_l & t_{l\leftarrow r}\\
        t_{r\leftarrow l} & r_r
    \end{pmatrix},
\end{equation}
where $r_l$ and $r_r$ are the reflection blocks, and $t_{i\leftarrow j}$ is the transmission block from lead $j$ to lead $i$. The total transmission from the left lead to the right lead is
\begin{equation}
    \mathcal{T}(E;g)
    =
    \operatorname{Tr}\!\left[
    t_{r\leftarrow l}^{\dagger}(E;g)
    t_{r\leftarrow l}(E;g)
    \right].
\end{equation}
At zero temperature and in linear response, the two-terminal conductance is
\begin{equation}
    G(E_F;g)
    =
    \frac{e^2}{h}\mathcal{T}(E_F;g),
    \label{eq:two-cond}
\end{equation}
where $E_F$ is the Fermi energy. The trace includes all propagating channels, including their spin degrees of freedom, so no additional spin-degeneracy factor is introduced. We compute the scattering matrix and conductance using \texttt{Kwant}~\cite{Groth_2014}. The resulting conductance provides a transport diagnostic of the auxiliary correlation Hamiltonian, which we compare with the bulk $\mathbb{Z}_2$ phase diagram.

\vbox{} \noindent \textbf{DATA AVAILABILITY}

\noindent The data set that supports the findings of the present study can be available from the corresponding authors via email upon reasonable request. 

\vbox{} \noindent \textbf{ACKNOWLEDGEMENTS}

\noindent \noindent Ashutosh acknowledges support
from the ANRF (Government of India) via Sanction
No. PDF/2025/001247, IISER Tirupati.Ankur thanks
IISER Tirupati for support, ANRF for start-up grant
ANRF/ECRG/2024/001172/PMS, and DST for grant
DST/NPNST/AM/2025/31.

\vbox{} \noindent \textbf{AUTHOR CONTRIBUTIONS}

\noindent All authors contributed equally to this work.

\vbox{} \noindent \textbf{COMPETING INTERESTS}

\noindent The authors declare no competing financial or non-financial interests.

\bibliography{reference}

\begin{thebibliography}{71}%
\makeatletter
\providecommand \@ifxundefined [1]{%
 \@ifx{#1\undefined}
}%
\providecommand \@ifnum [1]{%
 \ifnum #1\expandafter \@firstoftwo
 \else \expandafter \@secondoftwo
 \fi
}%
\providecommand \@ifx [1]{%
 \ifx #1\expandafter \@firstoftwo
 \else \expandafter \@secondoftwo
 \fi
}%
\providecommand \natexlab [1]{#1}%
\providecommand \enquote  [1]{``#1''}%
\providecommand \bibnamefont  [1]{#1}%
\providecommand \bibfnamefont [1]{#1}%
\providecommand \citenamefont [1]{#1}%
\providecommand \href@noop [0]{\@secondoftwo}%
\providecommand \href [0]{\begingroup \@sanitize@url \@href}%
\providecommand \@href[1]{\@@startlink{#1}\@@href}%
\providecommand \@@href[1]{\endgroup#1\@@endlink}%
\providecommand \@sanitize@url [0]{\catcode `\\12\catcode `\$12\catcode `\&12\catcode `\#12\catcode `\^12\catcode `\_12\catcode `\%12\relax}%
\providecommand \@@startlink[1]{}%
\providecommand \@@endlink[0]{}%
\providecommand \url  [0]{\begingroup\@sanitize@url \@url }%
\providecommand \@url [1]{\endgroup\@href {#1}{\urlprefix }}%
\providecommand \urlprefix  [0]{URL }%
\providecommand \Eprint [0]{\href }%
\providecommand \doibase [0]{https://doi.org/}%
\providecommand \selectlanguage [0]{\@gobble}%
\providecommand \bibinfo  [0]{\@secondoftwo}%
\providecommand \bibfield  [0]{\@secondoftwo}%
\providecommand \translation [1]{[#1]}%
\providecommand \BibitemOpen [0]{}%
\providecommand \bibitemStop [0]{}%
\providecommand \bibitemNoStop [0]{.\EOS\space}%
\providecommand \EOS [0]{\spacefactor3000\relax}%
\providecommand \BibitemShut  [1]{\csname bibitem#1\endcsname}%
\let\auto@bib@innerbib\@empty
\bibitem [{\citenamefont {Beekman}\ \emph {et~al.}(2019)\citenamefont {Beekman}, \citenamefont {Rademaker},\ and\ \citenamefont {van Wezel}}]{SciPostPhysLectNotes.11}%
  \BibitemOpen
  \bibfield  {author} {\bibinfo {author} {\bibfnamefont {A.}~\bibnamefont {Beekman}}, \bibinfo {author} {\bibfnamefont {L.}~\bibnamefont {Rademaker}},\ and\ \bibinfo {author} {\bibfnamefont {J.}~\bibnamefont {van Wezel}},\ }\bibfield  {title} {\bibinfo {title} {An introduction to spontaneous symmetry breaking},\ }\href {https://doi.org/10.21468/SciPostPhysLectNotes.11} {\bibfield  {journal} {\bibinfo  {journal} {SciPost Phys. Lect. Notes}\ ,\ \bibinfo {pages} {11}} (\bibinfo {year} {2019})}\BibitemShut {NoStop}%
\bibitem [{\citenamefont {Kundu}\ \emph {et~al.}(2023)\citenamefont {Kundu}, \citenamefont {Fertig},\ and\ \citenamefont {Kundu}}]{PhysRevB.107.L041402}%
  \BibitemOpen
  \bibfield  {author} {\bibinfo {author} {\bibfnamefont {R.}~\bibnamefont {Kundu}}, \bibinfo {author} {\bibfnamefont {H.~A.}\ \bibnamefont {Fertig}},\ and\ \bibinfo {author} {\bibfnamefont {A.}~\bibnamefont {Kundu}},\ }\bibfield  {title} {\bibinfo {title} {Broken symmetry and competing orders in weyl semimetal interfaces},\ }\href {https://doi.org/10.1103/PhysRevB.107.L041402} {\bibfield  {journal} {\bibinfo  {journal} {Phys. Rev. B}\ }\textbf {\bibinfo {volume} {107}},\ \bibinfo {pages} {L041402} (\bibinfo {year} {2023})}\BibitemShut {NoStop}%
\bibitem [{\citenamefont {Heyl}(2014)}]{PhysRevLett.113.205701}%
  \BibitemOpen
  \bibfield  {author} {\bibinfo {author} {\bibfnamefont {M.}~\bibnamefont {Heyl}},\ }\bibfield  {title} {\bibinfo {title} {Dynamical quantum phase transitions in systems with broken-symmetry phases},\ }\href {https://doi.org/10.1103/PhysRevLett.113.205701} {\bibfield  {journal} {\bibinfo  {journal} {Phys. Rev. Lett.}\ }\textbf {\bibinfo {volume} {113}},\ \bibinfo {pages} {205701} (\bibinfo {year} {2014})}\BibitemShut {NoStop}%
\bibitem [{\citenamefont {Gu}\ \emph {et~al.}(2025)\citenamefont {Gu}, \citenamefont {Wang},\ and\ \citenamefont {Wang}}]{3g6d-gn7b}%
  \BibitemOpen
  \bibfield  {author} {\bibinfo {author} {\bibfnamefont {D.}~\bibnamefont {Gu}}, \bibinfo {author} {\bibfnamefont {Z.}~\bibnamefont {Wang}},\ and\ \bibinfo {author} {\bibfnamefont {Z.}~\bibnamefont {Wang}},\ }\bibfield  {title} {\bibinfo {title} {Spontaneous symmetry breaking in open quantum systems: Strong, weak, and strong-to-weak},\ }\href {https://doi.org/10.1103/3g6d-gn7b} {\bibfield  {journal} {\bibinfo  {journal} {Phys. Rev. B}\ }\textbf {\bibinfo {volume} {112}},\ \bibinfo {pages} {245123} (\bibinfo {year} {2025})}\BibitemShut {NoStop}%
\bibitem [{\citenamefont {Qi}\ and\ \citenamefont {Zhang}(2011)}]{RevModPhys.83.1057}%
  \BibitemOpen
  \bibfield  {author} {\bibinfo {author} {\bibfnamefont {X.-L.}\ \bibnamefont {Qi}}\ and\ \bibinfo {author} {\bibfnamefont {S.-C.}\ \bibnamefont {Zhang}},\ }\bibfield  {title} {\bibinfo {title} {Topological insulators and superconductors},\ }\href {https://doi.org/10.1103/RevModPhys.83.1057} {\bibfield  {journal} {\bibinfo  {journal} {Rev. Mod. Phys.}\ }\textbf {\bibinfo {volume} {83}},\ \bibinfo {pages} {1057} (\bibinfo {year} {2011})}\BibitemShut {NoStop}%
\bibitem [{\citenamefont {Abdulla}\ \emph {et~al.}(2025)\citenamefont {Abdulla}, \citenamefont {Murthy},\ and\ \citenamefont {Das}}]{q4gd-b4vb}%
  \BibitemOpen
  \bibfield  {author} {\bibinfo {author} {\bibfnamefont {F.}~\bibnamefont {Abdulla}}, \bibinfo {author} {\bibfnamefont {G.}~\bibnamefont {Murthy}},\ and\ \bibinfo {author} {\bibfnamefont {A.}~\bibnamefont {Das}},\ }\bibfield  {title} {\bibinfo {title} {Internal symmetry protected stable topological semimetals in three dimensions},\ }\href {https://doi.org/10.1103/q4gd-b4vb} {\bibfield  {journal} {\bibinfo  {journal} {Phys. Rev. B}\ }\textbf {\bibinfo {volume} {112}},\ \bibinfo {pages} {205111} (\bibinfo {year} {2025})}\BibitemShut {NoStop}%
\bibitem [{\citenamefont {Das}\ \emph {et~al.}(2023)\citenamefont {Das}, \citenamefont {Cornfeld},\ and\ \citenamefont {Pujari}}]{PhysRevLett.130.186202}%
  \BibitemOpen
  \bibfield  {author} {\bibinfo {author} {\bibfnamefont {A.}~\bibnamefont {Das}}, \bibinfo {author} {\bibfnamefont {E.}~\bibnamefont {Cornfeld}},\ and\ \bibinfo {author} {\bibfnamefont {S.}~\bibnamefont {Pujari}},\ }\bibfield  {title} {\bibinfo {title} {Punctured-chern topological invariants for semimetallic band structures},\ }\href {https://doi.org/10.1103/PhysRevLett.130.186202} {\bibfield  {journal} {\bibinfo  {journal} {Phys. Rev. Lett.}\ }\textbf {\bibinfo {volume} {130}},\ \bibinfo {pages} {186202} (\bibinfo {year} {2023})}\BibitemShut {NoStop}%
\bibitem [{\citenamefont {Abdulla}\ \emph {et~al.}(2022)\citenamefont {Abdulla}, \citenamefont {Das}, \citenamefont {Rao},\ and\ \citenamefont {Murthy}}]{SciPostPhysCore.5.1.014}%
  \BibitemOpen
  \bibfield  {author} {\bibinfo {author} {\bibfnamefont {F.}~\bibnamefont {Abdulla}}, \bibinfo {author} {\bibfnamefont {A.}~\bibnamefont {Das}}, \bibinfo {author} {\bibfnamefont {S.}~\bibnamefont {Rao}},\ and\ \bibinfo {author} {\bibfnamefont {G.}~\bibnamefont {Murthy}},\ }\bibfield  {title} {\bibinfo {title} {Time-reversal-broken weyl semimetal in the hofstadter regime},\ }\href {https://doi.org/10.21468/SciPostPhysCore.5.1.014} {\bibfield  {journal} {\bibinfo  {journal} {SciPost Phys. Core}\ }\textbf {\bibinfo {volume} {5}},\ \bibinfo {pages} {014} (\bibinfo {year} {2022})}\BibitemShut {NoStop}%
\bibitem [{\citenamefont {Armitage}\ \emph {et~al.}(2018)\citenamefont {Armitage}, \citenamefont {Mele},\ and\ \citenamefont {Vishwanath}}]{RevModPhys.90.015001}%
  \BibitemOpen
  \bibfield  {author} {\bibinfo {author} {\bibfnamefont {N.~P.}\ \bibnamefont {Armitage}}, \bibinfo {author} {\bibfnamefont {E.~J.}\ \bibnamefont {Mele}},\ and\ \bibinfo {author} {\bibfnamefont {A.}~\bibnamefont {Vishwanath}},\ }\bibfield  {title} {\bibinfo {title} {Weyl and dirac semimetals in three-dimensional solids},\ }\href {https://doi.org/10.1103/RevModPhys.90.015001} {\bibfield  {journal} {\bibinfo  {journal} {Rev. Mod. Phys.}\ }\textbf {\bibinfo {volume} {90}},\ \bibinfo {pages} {015001} (\bibinfo {year} {2018})}\BibitemShut {NoStop}%
\bibitem [{\citenamefont {Yan}\ and\ \citenamefont {Felser}(2017)}]{Yan2017}%
  \BibitemOpen
  \bibfield  {author} {\bibinfo {author} {\bibfnamefont {B.}~\bibnamefont {Yan}}\ and\ \bibinfo {author} {\bibfnamefont {C.}~\bibnamefont {Felser}},\ }\bibfield  {title} {\bibinfo {title} {Topological materials: Weyl semimetals},\ }\href {https://doi.org/10.1146/annurev-conmatphys-031016-025458} {\bibfield  {journal} {\bibinfo  {journal} {Annu. Rev. Condens. Matter Phys.}\ }\textbf {\bibinfo {volume} {8}},\ \bibinfo {pages} {337} (\bibinfo {year} {2017})}\BibitemShut {NoStop}%
\bibitem [{\citenamefont {Gao}\ \emph {et~al.}(2019)\citenamefont {Gao}, \citenamefont {Venderbos}, \citenamefont {Kim},\ and\ \citenamefont {Rappe}}]{Gao2019}%
  \BibitemOpen
  \bibfield  {author} {\bibinfo {author} {\bibfnamefont {H.}~\bibnamefont {Gao}}, \bibinfo {author} {\bibfnamefont {J.~W.~F.}\ \bibnamefont {Venderbos}}, \bibinfo {author} {\bibfnamefont {Y.}~\bibnamefont {Kim}},\ and\ \bibinfo {author} {\bibfnamefont {A.~M.}\ \bibnamefont {Rappe}},\ }\bibfield  {title} {\bibinfo {title} {Topological semimetals from first principles},\ }\href {https://doi.org/10.1146/annurev-matsci-070218-010049} {\bibfield  {journal} {\bibinfo  {journal} {Annu. Rev. Mater. Res.}\ }\textbf {\bibinfo {volume} {49}},\ \bibinfo {pages} {153} (\bibinfo {year} {2019})}\BibitemShut {NoStop}%
\bibitem [{\citenamefont {Bernevig}\ and\ \citenamefont {Hughes}(2013)}]{bernevig2013topological}%
  \BibitemOpen
  \bibfield  {author} {\bibinfo {author} {\bibfnamefont {B.~A.}\ \bibnamefont {Bernevig}}\ and\ \bibinfo {author} {\bibfnamefont {T.~L.}\ \bibnamefont {Hughes}},\ }\href@noop {} {\emph {\bibinfo {title} {Topological insulators and topological superconductors}}}\ (\bibinfo  {publisher} {Princeton university press},\ \bibinfo {year} {2013})\BibitemShut {NoStop}%
\bibitem [{\citenamefont {Sato}\ and\ \citenamefont {Ando}(2017)}]{Sato2017}%
  \BibitemOpen
  \bibfield  {author} {\bibinfo {author} {\bibfnamefont {M.}~\bibnamefont {Sato}}\ and\ \bibinfo {author} {\bibfnamefont {Y.}~\bibnamefont {Ando}},\ }\bibfield  {title} {\bibinfo {title} {Topological superconductors: a review},\ }\href {https://doi.org/10.1088/1361-6633/aa6ac7} {\bibfield  {journal} {\bibinfo  {journal} {Rep. Prog. Phys.}\ }\textbf {\bibinfo {volume} {80}},\ \bibinfo {pages} {076501} (\bibinfo {year} {2017})}\BibitemShut {NoStop}%
\bibitem [{\citenamefont {Kitamura}\ and\ \citenamefont {Aoki}(2022)}]{Kitamura2022}%
  \BibitemOpen
  \bibfield  {author} {\bibinfo {author} {\bibfnamefont {S.}~\bibnamefont {Kitamura}}\ and\ \bibinfo {author} {\bibfnamefont {H.}~\bibnamefont {Aoki}},\ }\bibfield  {title} {\bibinfo {title} {Floquet topological superconductivity induced by chiral many-body interaction},\ }\href {https://doi.org/10.1038/s42005-022-00936-w} {\bibfield  {journal} {\bibinfo  {journal} {Commun. Phys.}\ }\textbf {\bibinfo {volume} {5}},\ \bibinfo {pages} {174} (\bibinfo {year} {2022})}\BibitemShut {NoStop}%
\bibitem [{\citenamefont {Wang}\ \emph {et~al.}(2023)\citenamefont {Wang}, \citenamefont {Wang}, \citenamefont {Sun}, \citenamefont {Chen},\ and\ \citenamefont {Xu}}]{PhysRevB.107.L121407}%
  \BibitemOpen
  \bibfield  {author} {\bibinfo {author} {\bibfnamefont {Z.-M.}\ \bibnamefont {Wang}}, \bibinfo {author} {\bibfnamefont {R.}~\bibnamefont {Wang}}, \bibinfo {author} {\bibfnamefont {J.-H.}\ \bibnamefont {Sun}}, \bibinfo {author} {\bibfnamefont {T.-Y.}\ \bibnamefont {Chen}},\ and\ \bibinfo {author} {\bibfnamefont {D.-H.}\ \bibnamefont {Xu}},\ }\bibfield  {title} {\bibinfo {title} {Floquet weyl semimetal phases in light-irradiated higher-order topological dirac semimetals},\ }\href {https://doi.org/10.1103/PhysRevB.107.L121407} {\bibfield  {journal} {\bibinfo  {journal} {Phys. Rev. B}\ }\textbf {\bibinfo {volume} {107}},\ \bibinfo {pages} {L121407} (\bibinfo {year} {2023})}\BibitemShut {NoStop}%
\bibitem [{\citenamefont {Dubey}\ \emph {et~al.}(2025)\citenamefont {Dubey}, \citenamefont {Kundu},\ and\ \citenamefont {Kundu}}]{PhysRevB.111.035431}%
  \BibitemOpen
  \bibfield  {author} {\bibinfo {author} {\bibfnamefont {A.}~\bibnamefont {Dubey}}, \bibinfo {author} {\bibfnamefont {R.}~\bibnamefont {Kundu}},\ and\ \bibinfo {author} {\bibfnamefont {A.}~\bibnamefont {Kundu}},\ }\bibfield  {title} {\bibinfo {title} {Time-resolved arpes and optical transport properties of irradiated twisted bilayer graphene in a steady state},\ }\href {https://doi.org/10.1103/PhysRevB.111.035431} {\bibfield  {journal} {\bibinfo  {journal} {Phys. Rev. B}\ }\textbf {\bibinfo {volume} {111}},\ \bibinfo {pages} {035431} (\bibinfo {year} {2025})}\BibitemShut {NoStop}%
\bibitem [{\citenamefont {Kundu}\ \emph {et~al.}(2014)\citenamefont {Kundu}, \citenamefont {Fertig},\ and\ \citenamefont {Seradjeh}}]{PhysRevLett.113.236803}%
  \BibitemOpen
  \bibfield  {author} {\bibinfo {author} {\bibfnamefont {A.}~\bibnamefont {Kundu}}, \bibinfo {author} {\bibfnamefont {H.~A.}\ \bibnamefont {Fertig}},\ and\ \bibinfo {author} {\bibfnamefont {B.}~\bibnamefont {Seradjeh}},\ }\bibfield  {title} {\bibinfo {title} {Effective theory of floquet topological transitions},\ }\href {https://doi.org/10.1103/PhysRevLett.113.236803} {\bibfield  {journal} {\bibinfo  {journal} {Phys. Rev. Lett.}\ }\textbf {\bibinfo {volume} {113}},\ \bibinfo {pages} {236803} (\bibinfo {year} {2014})}\BibitemShut {NoStop}%
\bibitem [{\citenamefont {Dubey}\ \emph {et~al.}(2026)\citenamefont {Dubey}, \citenamefont {Pratihar}, \citenamefont {Sen},\ and\ \citenamefont {Kundu}}]{dubey2026topologicalsignaturesquenchdynamics}%
  \BibitemOpen
  \bibfield  {author} {\bibinfo {author} {\bibfnamefont {A.}~\bibnamefont {Dubey}}, \bibinfo {author} {\bibfnamefont {H.}~\bibnamefont {Pratihar}}, \bibinfo {author} {\bibfnamefont {D.}~\bibnamefont {Sen}},\ and\ \bibinfo {author} {\bibfnamefont {A.}~\bibnamefont {Kundu}},\ }\href {https://arxiv.org/abs/2608.27588} {\bibinfo {title} {Topological signatures in the quench dynamics of periodically driven quantum systems}} (\bibinfo {year} {2026}),\ \Eprint {https://arxiv.org/abs/2608.27588} {arXiv:2608.27588 [cond-mat.mes-hall]} \BibitemShut {NoStop}%
\bibitem [{\citenamefont {Azizi}(2025)}]{Azizi2025}%
  \BibitemOpen
  \bibfield  {author} {\bibinfo {author} {\bibfnamefont {F.}~\bibnamefont {Azizi}},\ }\bibfield  {title} {\bibinfo {title} {Strain engineering of topological transitions in 2d materials: a multi-band approach},\ }\href {https://doi.org/10.1038/s41598-025-28092-5} {\bibfield  {journal} {\bibinfo  {journal} {Sci. Rep.}\ }\textbf {\bibinfo {volume} {15}},\ \bibinfo {pages} {44367} (\bibinfo {year} {2025})}\BibitemShut {NoStop}%
\bibitem [{\citenamefont {Lukmanova}\ \emph {et~al.}(2025)\citenamefont {Lukmanova}, \citenamefont {Cohn}, \citenamefont {Minakova},\ and\ \citenamefont {Zaitsev-Zotov}}]{dccn-czmz}%
  \BibitemOpen
  \bibfield  {author} {\bibinfo {author} {\bibfnamefont {R.~M.}\ \bibnamefont {Lukmanova}}, \bibinfo {author} {\bibfnamefont {I.~A.}\ \bibnamefont {Cohn}}, \bibinfo {author} {\bibfnamefont {V.~E.}\ \bibnamefont {Minakova}},\ and\ \bibinfo {author} {\bibfnamefont {S.~V.}\ \bibnamefont {Zaitsev-Zotov}},\ }\bibfield  {title} {\bibinfo {title} {Strain-induced superconducting proximity effect in the topological insulator ${\mathrm{tase}}_{3}$},\ }\href {https://doi.org/10.1103/dccn-czmz} {\bibfield  {journal} {\bibinfo  {journal} {Phys. Rev. B}\ }\textbf {\bibinfo {volume} {111}},\ \bibinfo {pages} {224510} (\bibinfo {year} {2025})}\BibitemShut {NoStop}%
\bibitem [{\citenamefont {Cortijo}\ \emph {et~al.}(2016)\citenamefont {Cortijo}, \citenamefont {Kharzeev}, \citenamefont {Landsteiner},\ and\ \citenamefont {Vozmediano}}]{PhysRevB.94.241405}%
  \BibitemOpen
  \bibfield  {author} {\bibinfo {author} {\bibfnamefont {A.}~\bibnamefont {Cortijo}}, \bibinfo {author} {\bibfnamefont {D.}~\bibnamefont {Kharzeev}}, \bibinfo {author} {\bibfnamefont {K.}~\bibnamefont {Landsteiner}},\ and\ \bibinfo {author} {\bibfnamefont {M.~A.~H.}\ \bibnamefont {Vozmediano}},\ }\bibfield  {title} {\bibinfo {title} {Strain-induced chiral magnetic effect in weyl semimetals},\ }\href {https://doi.org/10.1103/PhysRevB.94.241405} {\bibfield  {journal} {\bibinfo  {journal} {Phys. Rev. B}\ }\textbf {\bibinfo {volume} {94}},\ \bibinfo {pages} {241405(R)} (\bibinfo {year} {2016})}\BibitemShut {NoStop}%
\bibitem [{\citenamefont {Zhang}\ \emph {et~al.}(2021)\citenamefont {Zhang}, \citenamefont {Zhang},\ and\ \citenamefont {Liu}}]{PhysRevResearch.3.013229}%
  \BibitemOpen
  \bibfield  {author} {\bibinfo {author} {\bibfnamefont {L.}~\bibnamefont {Zhang}}, \bibinfo {author} {\bibfnamefont {L.}~\bibnamefont {Zhang}},\ and\ \bibinfo {author} {\bibfnamefont {X.-J.}\ \bibnamefont {Liu}},\ }\bibfield  {title} {\bibinfo {title} {Quench-induced dynamical topology under dynamical noise},\ }\href {https://doi.org/10.1103/PhysRevResearch.3.013229} {\bibfield  {journal} {\bibinfo  {journal} {Phys. Rev. Res.}\ }\textbf {\bibinfo {volume} {3}},\ \bibinfo {pages} {013229} (\bibinfo {year} {2021})}\BibitemShut {NoStop}%
\bibitem [{\citenamefont {Xu}\ \emph {et~al.}(2017)\citenamefont {Xu}, \citenamefont {Chiu}, \citenamefont {Miao}, \citenamefont {He}, \citenamefont {Alpichshev}, \citenamefont {Kapitulnik}, \citenamefont {Biswas},\ and\ \citenamefont {Wray}}]{Xu2017}%
  \BibitemOpen
  \bibfield  {author} {\bibinfo {author} {\bibfnamefont {Y.}~\bibnamefont {Xu}}, \bibinfo {author} {\bibfnamefont {J.}~\bibnamefont {Chiu}}, \bibinfo {author} {\bibfnamefont {L.}~\bibnamefont {Miao}}, \bibinfo {author} {\bibfnamefont {H.}~\bibnamefont {He}}, \bibinfo {author} {\bibfnamefont {Z.}~\bibnamefont {Alpichshev}}, \bibinfo {author} {\bibfnamefont {A.}~\bibnamefont {Kapitulnik}}, \bibinfo {author} {\bibfnamefont {R.~R.}\ \bibnamefont {Biswas}},\ and\ \bibinfo {author} {\bibfnamefont {L.~A.}\ \bibnamefont {Wray}},\ }\bibfield  {title} {\bibinfo {title} {Disorder enabled band structure engineering of a topological insulator surface},\ }\href {https://doi.org/10.1038/ncomms14081} {\bibfield  {journal} {\bibinfo  {journal} {Nat. Commun.}\ }\textbf {\bibinfo {volume} {8}},\ \bibinfo {pages} {14081} (\bibinfo {year} {2017})}\BibitemShut {NoStop}%
\bibitem [{\citenamefont {Antonius}\ and\ \citenamefont {Louie}(2016)}]{Antonius2016}%
  \BibitemOpen
  \bibfield  {author} {\bibinfo {author} {\bibfnamefont {G.}~\bibnamefont {Antonius}}\ and\ \bibinfo {author} {\bibfnamefont {S.~G.}\ \bibnamefont {Louie}},\ }\bibfield  {title} {\bibinfo {title} {Temperature-induced topological phase transitions: Promoted versus suppressed nontrivial topology},\ }\href {https://doi.org/10.1103/PhysRevLett.117.246401} {\bibfield  {journal} {\bibinfo  {journal} {Phys. Rev. Lett.}\ }\textbf {\bibinfo {volume} {117}},\ \bibinfo {pages} {246401} (\bibinfo {year} {2016})}\BibitemShut {NoStop}%
\bibitem [{\citenamefont {Ryu}\ \emph {et~al.}(2010)\citenamefont {Ryu}, \citenamefont {Schnyder}, \citenamefont {Furusaki},\ and\ \citenamefont {Ludwig}}]{Ryu2010}%
  \BibitemOpen
  \bibfield  {author} {\bibinfo {author} {\bibfnamefont {S.}~\bibnamefont {Ryu}}, \bibinfo {author} {\bibfnamefont {A.~P.}\ \bibnamefont {Schnyder}}, \bibinfo {author} {\bibfnamefont {A.}~\bibnamefont {Furusaki}},\ and\ \bibinfo {author} {\bibfnamefont {A.~W.~W.}\ \bibnamefont {Ludwig}},\ }\bibfield  {title} {\bibinfo {title} {Topological insulators and superconductors: tenfold way and dimensional hierarchy},\ }\href {https://doi.org/10.1088/1367-2630/12/6/065010} {\bibfield  {journal} {\bibinfo  {journal} {New J. Phys.}\ }\textbf {\bibinfo {volume} {12}},\ \bibinfo {pages} {065010} (\bibinfo {year} {2010})}\BibitemShut {NoStop}%
\bibitem [{\citenamefont {Haldane}(1988)}]{PhysRevLett.61.2015}%
  \BibitemOpen
  \bibfield  {author} {\bibinfo {author} {\bibfnamefont {F.~D.~M.}\ \bibnamefont {Haldane}},\ }\bibfield  {title} {\bibinfo {title} {Model for a quantum hall effect without landau levels: Condensed-matter realization of the "parity anomaly"},\ }\href {https://doi.org/10.1103/PhysRevLett.61.2015} {\bibfield  {journal} {\bibinfo  {journal} {Phys. Rev. Lett.}\ }\textbf {\bibinfo {volume} {61}},\ \bibinfo {pages} {2015} (\bibinfo {year} {1988})}\BibitemShut {NoStop}%
\bibitem [{\citenamefont {Kane}\ and\ \citenamefont {Mele}(2005)}]{PhysRevLett.95.146802}%
  \BibitemOpen
  \bibfield  {author} {\bibinfo {author} {\bibfnamefont {C.~L.}\ \bibnamefont {Kane}}\ and\ \bibinfo {author} {\bibfnamefont {E.~J.}\ \bibnamefont {Mele}},\ }\bibfield  {title} {\bibinfo {title} {${Z}_{2}$ topological order and the quantum spin hall effect},\ }\href {https://doi.org/10.1103/PhysRevLett.95.146802} {\bibfield  {journal} {\bibinfo  {journal} {Phys. Rev. Lett.}\ }\textbf {\bibinfo {volume} {95}},\ \bibinfo {pages} {146802} (\bibinfo {year} {2005})}\BibitemShut {NoStop}%
\bibitem [{\citenamefont {Goldman}\ and\ \citenamefont {Ozawa}(2024)}]{Goldman2024}%
  \BibitemOpen
  \bibfield  {author} {\bibinfo {author} {\bibfnamefont {N.}~\bibnamefont {Goldman}}\ and\ \bibinfo {author} {\bibfnamefont {T.}~\bibnamefont {Ozawa}},\ }\bibfield  {title} {\bibinfo {title} {Relating the hall conductivity to the many-body chern number using fermi's golden rule and kramers--kronig relations},\ }\href {https://doi.org/10.5802/crphys.191} {\bibfield  {journal} {\bibinfo  {journal} {Comptes Rendus Physique}\ }\textbf {\bibinfo {volume} {25}},\ \bibinfo {pages} {289} (\bibinfo {year} {2024})}\BibitemShut {NoStop}%
\bibitem [{\citenamefont {Cooper}\ \emph {et~al.}(2019)\citenamefont {Cooper}, \citenamefont {Dalibard},\ and\ \citenamefont {Spielman}}]{RevModPhys.91.015005}%
  \BibitemOpen
  \bibfield  {author} {\bibinfo {author} {\bibfnamefont {N.~R.}\ \bibnamefont {Cooper}}, \bibinfo {author} {\bibfnamefont {J.}~\bibnamefont {Dalibard}},\ and\ \bibinfo {author} {\bibfnamefont {I.~B.}\ \bibnamefont {Spielman}},\ }\bibfield  {title} {\bibinfo {title} {Topological bands for ultracold atoms},\ }\href {https://doi.org/10.1103/RevModPhys.91.015005} {\bibfield  {journal} {\bibinfo  {journal} {Rev. Mod. Phys.}\ }\textbf {\bibinfo {volume} {91}},\ \bibinfo {pages} {015005} (\bibinfo {year} {2019})}\BibitemShut {NoStop}%
\bibitem [{\citenamefont {Wang}\ \emph {et~al.}(2026{\natexlab{a}})\citenamefont {Wang}, \citenamefont {Lin}, \citenamefont {Zhang},\ and\ \citenamefont {Zhang}}]{Wang2026}%
  \BibitemOpen
  \bibfield  {author} {\bibinfo {author} {\bibfnamefont {B.-B.}\ \bibnamefont {Wang}}, \bibinfo {author} {\bibfnamefont {X.-D.}\ \bibnamefont {Lin}}, \bibinfo {author} {\bibfnamefont {J.}~\bibnamefont {Zhang}},\ and\ \bibinfo {author} {\bibfnamefont {L.}~\bibnamefont {Zhang}},\ }\bibfield  {title} {\bibinfo {title} {Ultracold atomic lattice systems for simulating topological phases: A review},\ }\href {https://doi.org/10.1016/j.qrl.2026.06.003} {\bibfield  {journal} {\bibinfo  {journal} {Quantum Review Letters}\ }\textbf {\bibinfo {volume} {2}},\ \bibinfo {pages} {117} (\bibinfo {year} {2026}{\natexlab{a}})}\BibitemShut {NoStop}%
\bibitem [{\citenamefont {B\'eri}\ and\ \citenamefont {Cooper}(2011)}]{PhysRevLett.107.145301}%
  \BibitemOpen
  \bibfield  {author} {\bibinfo {author} {\bibfnamefont {B.}~\bibnamefont {B\'eri}}\ and\ \bibinfo {author} {\bibfnamefont {N.~R.}\ \bibnamefont {Cooper}},\ }\bibfield  {title} {\bibinfo {title} {${\mathbb{z}}_{2}$ topological insulators in ultracold atomic gases},\ }\href {https://doi.org/10.1103/PhysRevLett.107.145301} {\bibfield  {journal} {\bibinfo  {journal} {Phys. Rev. Lett.}\ }\textbf {\bibinfo {volume} {107}},\ \bibinfo {pages} {145301} (\bibinfo {year} {2011})}\BibitemShut {NoStop}%
\bibitem [{\citenamefont {Ozawa}\ \emph {et~al.}(2019)\citenamefont {Ozawa}, \citenamefont {Price}, \citenamefont {Amo}, \citenamefont {Goldman}, \citenamefont {Hafezi}, \citenamefont {Lu}, \citenamefont {Rechtsman}, \citenamefont {Schuster}, \citenamefont {Simon}, \citenamefont {Zilberberg},\ and\ \citenamefont {Carusotto}}]{RevModPhys.91.015006}%
  \BibitemOpen
  \bibfield  {author} {\bibinfo {author} {\bibfnamefont {T.}~\bibnamefont {Ozawa}}, \bibinfo {author} {\bibfnamefont {H.~M.}\ \bibnamefont {Price}}, \bibinfo {author} {\bibfnamefont {A.}~\bibnamefont {Amo}}, \bibinfo {author} {\bibfnamefont {N.}~\bibnamefont {Goldman}}, \bibinfo {author} {\bibfnamefont {M.}~\bibnamefont {Hafezi}}, \bibinfo {author} {\bibfnamefont {L.}~\bibnamefont {Lu}}, \bibinfo {author} {\bibfnamefont {M.~C.}\ \bibnamefont {Rechtsman}}, \bibinfo {author} {\bibfnamefont {D.}~\bibnamefont {Schuster}}, \bibinfo {author} {\bibfnamefont {J.}~\bibnamefont {Simon}}, \bibinfo {author} {\bibfnamefont {O.}~\bibnamefont {Zilberberg}},\ and\ \bibinfo {author} {\bibfnamefont {I.}~\bibnamefont {Carusotto}},\ }\bibfield  {title} {\bibinfo {title} {Topological photonics},\ }\href {https://doi.org/10.1103/RevModPhys.91.015006} {\bibfield  {journal} {\bibinfo  {journal} {Rev. Mod. Phys.}\ }\textbf {\bibinfo {volume} {91}},\ \bibinfo {pages} {015006} (\bibinfo {year} {2019})}\BibitemShut {NoStop}%
\bibitem [{\citenamefont {Lan}\ \emph {et~al.}(2022)\citenamefont {Lan}, \citenamefont {Chen}, \citenamefont {Gao}, \citenamefont {Zhang},\ and\ \citenamefont {Sha}}]{Lan2022}%
  \BibitemOpen
  \bibfield  {author} {\bibinfo {author} {\bibfnamefont {Z.}~\bibnamefont {Lan}}, \bibinfo {author} {\bibfnamefont {M.~L.~N.}\ \bibnamefont {Chen}}, \bibinfo {author} {\bibfnamefont {F.}~\bibnamefont {Gao}}, \bibinfo {author} {\bibfnamefont {S.}~\bibnamefont {Zhang}},\ and\ \bibinfo {author} {\bibfnamefont {W.~E.~I.}\ \bibnamefont {Sha}},\ }\bibfield  {title} {\bibinfo {title} {A brief review of topological photonics in one, two, and three dimensions},\ }\href {https://doi.org/10.1016/j.revip.2022.100076} {\bibfield  {journal} {\bibinfo  {journal} {Rev. Phys.}\ }\textbf {\bibinfo {volume} {9}},\ \bibinfo {pages} {100076} (\bibinfo {year} {2022})}\BibitemShut {NoStop}%
\bibitem [{\citenamefont {Khanikaev}\ and\ \citenamefont {Al{\`u}}(2024)}]{Khanikaev2024}%
  \BibitemOpen
  \bibfield  {author} {\bibinfo {author} {\bibfnamefont {A.~B.}\ \bibnamefont {Khanikaev}}\ and\ \bibinfo {author} {\bibfnamefont {A.}~\bibnamefont {Al{\`u}}},\ }\bibfield  {title} {\bibinfo {title} {Topological photonics: robustness and beyond},\ }\href {https://doi.org/10.1038/s41467-024-45194-2} {\bibfield  {journal} {\bibinfo  {journal} {Nat. Commun.}\ }\textbf {\bibinfo {volume} {15}},\ \bibinfo {pages} {931} (\bibinfo {year} {2024})}\BibitemShut {NoStop}%
\bibitem [{\citenamefont {Sridhar}\ \emph {et~al.}(2024)\citenamefont {Sridhar}, \citenamefont {Ghosh}, \citenamefont {Srinivasan}, \citenamefont {Miller},\ and\ \citenamefont {Dutt}}]{Sridhar2024}%
  \BibitemOpen
  \bibfield  {author} {\bibinfo {author} {\bibfnamefont {S.~K.}\ \bibnamefont {Sridhar}}, \bibinfo {author} {\bibfnamefont {S.}~\bibnamefont {Ghosh}}, \bibinfo {author} {\bibfnamefont {D.}~\bibnamefont {Srinivasan}}, \bibinfo {author} {\bibfnamefont {A.~R.}\ \bibnamefont {Miller}},\ and\ \bibinfo {author} {\bibfnamefont {A.}~\bibnamefont {Dutt}},\ }\bibfield  {title} {\bibinfo {title} {Quantized topological pumping in floquet synthetic dimensions with a driven dissipative photonic molecule},\ }\href {https://doi.org/10.1038/s41567-024-02413-3} {\bibfield  {journal} {\bibinfo  {journal} {Nat. Phys.}\ }\textbf {\bibinfo {volume} {20}},\ \bibinfo {pages} {843} (\bibinfo {year} {2024})}\BibitemShut {NoStop}%
\bibitem [{\citenamefont {Martin}\ \emph {et~al.}(2017)\citenamefont {Martin}, \citenamefont {Refael},\ and\ \citenamefont {Halperin}}]{PhysRevX.7.041008}%
  \BibitemOpen
  \bibfield  {author} {\bibinfo {author} {\bibfnamefont {I.}~\bibnamefont {Martin}}, \bibinfo {author} {\bibfnamefont {G.}~\bibnamefont {Refael}},\ and\ \bibinfo {author} {\bibfnamefont {B.}~\bibnamefont {Halperin}},\ }\bibfield  {title} {\bibinfo {title} {Topological frequency conversion in strongly driven quantum systems},\ }\href {https://doi.org/10.1103/PhysRevX.7.041008} {\bibfield  {journal} {\bibinfo  {journal} {Phys. Rev. X}\ }\textbf {\bibinfo {volume} {7}},\ \bibinfo {pages} {041008} (\bibinfo {year} {2017})}\BibitemShut {NoStop}%
\bibitem [{\citenamefont {Vinjamuri}\ \emph {et~al.}(2026)\citenamefont {Vinjamuri}, \citenamefont {Dubey},\ and\ \citenamefont {Das}}]{vinjamuri2026simulatingmultinodeweylsemimetals}%
  \BibitemOpen
  \bibfield  {author} {\bibinfo {author} {\bibfnamefont {G.}~\bibnamefont {Vinjamuri}}, \bibinfo {author} {\bibfnamefont {A.}~\bibnamefont {Dubey}},\ and\ \bibinfo {author} {\bibfnamefont {A.}~\bibnamefont {Das}},\ }\href {https://arxiv.org/abs/2606.20378} {\bibinfo {title} {Simulating multi-node weyl semimetals in a mixed floquet lattice}} (\bibinfo {year} {2026}),\ \Eprint {https://arxiv.org/abs/2606.20378} {arXiv:2606.20378 [cond-mat.mes-hall]} \BibitemShut {NoStop}%
\bibitem [{\citenamefont {Youssefi}\ \emph {et~al.}(2022)\citenamefont {Youssefi}, \citenamefont {Kono}, \citenamefont {Bancora}, \citenamefont {Chegnizadeh}, \citenamefont {Pan}, \citenamefont {Vovk},\ and\ \citenamefont {Kippenberg}}]{Youssefi2022}%
  \BibitemOpen
  \bibfield  {author} {\bibinfo {author} {\bibfnamefont {A.}~\bibnamefont {Youssefi}}, \bibinfo {author} {\bibfnamefont {S.}~\bibnamefont {Kono}}, \bibinfo {author} {\bibfnamefont {A.}~\bibnamefont {Bancora}}, \bibinfo {author} {\bibfnamefont {M.}~\bibnamefont {Chegnizadeh}}, \bibinfo {author} {\bibfnamefont {J.}~\bibnamefont {Pan}}, \bibinfo {author} {\bibfnamefont {T.}~\bibnamefont {Vovk}},\ and\ \bibinfo {author} {\bibfnamefont {T.~J.}\ \bibnamefont {Kippenberg}},\ }\bibfield  {title} {\bibinfo {title} {Topological lattices realized in superconducting circuit optomechanics},\ }\href {https://doi.org/10.1038/s41586-022-05367-9} {\bibfield  {journal} {\bibinfo  {journal} {Nature}\ }\textbf {\bibinfo {volume} {612}},\ \bibinfo {pages} {666} (\bibinfo {year} {2022})}\BibitemShut {NoStop}%
\bibitem [{\citenamefont {Cai}\ \emph {et~al.}(2019)\citenamefont {Cai}, \citenamefont {Han}, \citenamefont {Mei}, \citenamefont {Xu}, \citenamefont {Ma}, \citenamefont {Li}, \citenamefont {Wang}, \citenamefont {Song}, \citenamefont {Xue}, \citenamefont {Yin}, \citenamefont {Jia},\ and\ \citenamefont {Sun}}]{PhysRevLett.123.080501}%
  \BibitemOpen
  \bibfield  {author} {\bibinfo {author} {\bibfnamefont {W.}~\bibnamefont {Cai}}, \bibinfo {author} {\bibfnamefont {J.}~\bibnamefont {Han}}, \bibinfo {author} {\bibfnamefont {F.}~\bibnamefont {Mei}}, \bibinfo {author} {\bibfnamefont {Y.}~\bibnamefont {Xu}}, \bibinfo {author} {\bibfnamefont {Y.}~\bibnamefont {Ma}}, \bibinfo {author} {\bibfnamefont {X.}~\bibnamefont {Li}}, \bibinfo {author} {\bibfnamefont {H.}~\bibnamefont {Wang}}, \bibinfo {author} {\bibfnamefont {Y.~P.}\ \bibnamefont {Song}}, \bibinfo {author} {\bibfnamefont {Z.-Y.}\ \bibnamefont {Xue}}, \bibinfo {author} {\bibfnamefont {Z.-q.}\ \bibnamefont {Yin}}, \bibinfo {author} {\bibfnamefont {S.}~\bibnamefont {Jia}},\ and\ \bibinfo {author} {\bibfnamefont {L.}~\bibnamefont {Sun}},\ }\bibfield  {title} {\bibinfo {title} {Observation of topological magnon insulator states in a superconducting circuit},\ }\href {https://doi.org/10.1103/PhysRevLett.123.080501} {\bibfield  {journal} {\bibinfo  {journal} {Phys. Rev. Lett.}\ }\textbf {\bibinfo {volume}
  {123}},\ \bibinfo {pages} {080501} (\bibinfo {year} {2019})}\BibitemShut {NoStop}%
\bibitem [{\citenamefont {Peyruchat}\ \emph {et~al.}(2024)\citenamefont {Peyruchat}, \citenamefont {Rodriguez}, \citenamefont {Smirr}, \citenamefont {Leone},\ and\ \citenamefont {Girit}}]{PhysRevX.14.041041}%
  \BibitemOpen
  \bibfield  {author} {\bibinfo {author} {\bibfnamefont {L.}~\bibnamefont {Peyruchat}}, \bibinfo {author} {\bibfnamefont {R.~H.}\ \bibnamefont {Rodriguez}}, \bibinfo {author} {\bibfnamefont {J.-L.}\ \bibnamefont {Smirr}}, \bibinfo {author} {\bibfnamefont {R.}~\bibnamefont {Leone}},\ and\ \bibinfo {author} {\bibfnamefont {i.~m. c.~O.}\ \bibnamefont {Girit}},\ }\bibfield  {title} {\bibinfo {title} {Spectral signatures of nontrivial topology in a superconducting circuit},\ }\href {https://doi.org/10.1103/PhysRevX.14.041041} {\bibfield  {journal} {\bibinfo  {journal} {Phys. Rev. X}\ }\textbf {\bibinfo {volume} {14}},\ \bibinfo {pages} {041041} (\bibinfo {year} {2024})}\BibitemShut {NoStop}%
\bibitem [{\citenamefont {Gurevich}\ \emph {et~al.}(2026)\citenamefont {Gurevich}, \citenamefont {Xie}, \citenamefont {Yarmohammadi},\ and\ \citenamefont {Kolodrubetz}}]{w2w5-n3lp}%
  \BibitemOpen
  \bibfield  {author} {\bibinfo {author} {\bibfnamefont {B.}~\bibnamefont {Gurevich}}, \bibinfo {author} {\bibfnamefont {W.}~\bibnamefont {Xie}}, \bibinfo {author} {\bibfnamefont {M.}~\bibnamefont {Yarmohammadi}},\ and\ \bibinfo {author} {\bibfnamefont {M.~H.}\ \bibnamefont {Kolodrubetz}},\ }\bibfield  {title} {\bibinfo {title} {Robust qubit interactions mediated by photonic topological edge states},\ }\href {https://doi.org/10.1103/w2w5-n3lp} {\bibfield  {journal} {\bibinfo  {journal} {Phys. Rev. B}\ }\textbf {\bibinfo {volume} {113}},\ \bibinfo {pages} {045150} (\bibinfo {year} {2026})}\BibitemShut {NoStop}%
\bibitem [{\citenamefont {Wang}(2020)}]{Wang2020}%
  \BibitemOpen
  \bibfield  {author} {\bibinfo {author} {\bibfnamefont {D.-S.}\ \bibnamefont {Wang}},\ }\bibfield  {title} {\bibinfo {title} {Classes of topological qubits from low-dimensional quantum spin systems},\ }\href {https://doi.org/10.1016/j.aop.2019.168015} {\bibfield  {journal} {\bibinfo  {journal} {Ann. Phys.}\ }\textbf {\bibinfo {volume} {412}},\ \bibinfo {pages} {168015} (\bibinfo {year} {2020})}\BibitemShut {NoStop}%
\bibitem [{\citenamefont {Niedermeier}\ \emph {et~al.}(2024)\citenamefont {Niedermeier}, \citenamefont {Nairn}, \citenamefont {Flindt},\ and\ \citenamefont {Lado}}]{PhysRevResearch.6.043288}%
  \BibitemOpen
  \bibfield  {author} {\bibinfo {author} {\bibfnamefont {M.}~\bibnamefont {Niedermeier}}, \bibinfo {author} {\bibfnamefont {M.}~\bibnamefont {Nairn}}, \bibinfo {author} {\bibfnamefont {C.}~\bibnamefont {Flindt}},\ and\ \bibinfo {author} {\bibfnamefont {J.~L.}\ \bibnamefont {Lado}},\ }\bibfield  {title} {\bibinfo {title} {Quantum computing topological invariants of two-dimensional quantum matter},\ }\href {https://doi.org/10.1103/PhysRevResearch.6.043288} {\bibfield  {journal} {\bibinfo  {journal} {Phys. Rev. Res.}\ }\textbf {\bibinfo {volume} {6}},\ \bibinfo {pages} {043288} (\bibinfo {year} {2024})}\BibitemShut {NoStop}%
\bibitem [{\citenamefont {Rau}\ and\ \citenamefont {Wendell}(2002)}]{PhysRevLett.89.220405}%
  \BibitemOpen
  \bibfield  {author} {\bibinfo {author} {\bibfnamefont {A.~R.~P.}\ \bibnamefont {Rau}}\ and\ \bibinfo {author} {\bibfnamefont {R.~A.}\ \bibnamefont {Wendell}},\ }\bibfield  {title} {\bibinfo {title} {Embedding dissipation and decoherence in unitary evolution schemes},\ }\href {https://doi.org/10.1103/PhysRevLett.89.220405} {\bibfield  {journal} {\bibinfo  {journal} {Phys. Rev. Lett.}\ }\textbf {\bibinfo {volume} {89}},\ \bibinfo {pages} {220405} (\bibinfo {year} {2002})}\BibitemShut {NoStop}%
\bibitem [{\citenamefont {Furuta}(2001)}]{PhysRevA.64.042110}%
  \BibitemOpen
  \bibfield  {author} {\bibinfo {author} {\bibfnamefont {S.}~\bibnamefont {Furuta}},\ }\bibfield  {title} {\bibinfo {title} {Measurement-induced decoherence in a simple quantum system},\ }\href {https://doi.org/10.1103/PhysRevA.64.042110} {\bibfield  {journal} {\bibinfo  {journal} {Phys. Rev. A}\ }\textbf {\bibinfo {volume} {64}},\ \bibinfo {pages} {042110} (\bibinfo {year} {2001})}\BibitemShut {NoStop}%
\bibitem [{\citenamefont {Rotter}\ and\ \citenamefont {Bird}(2015)}]{Rotter2015}%
  \BibitemOpen
  \bibfield  {author} {\bibinfo {author} {\bibfnamefont {I.}~\bibnamefont {Rotter}}\ and\ \bibinfo {author} {\bibfnamefont {J.~P.}\ \bibnamefont {Bird}},\ }\bibfield  {title} {\bibinfo {title} {A review of progress in the physics of open quantum systems: theory and experiment},\ }\href {https://doi.org/10.1088/0034-4885/78/11/114001} {\bibfield  {journal} {\bibinfo  {journal} {Rep. Prog. Phys.}\ }\textbf {\bibinfo {volume} {78}},\ \bibinfo {pages} {114001} (\bibinfo {year} {2015})}\BibitemShut {NoStop}%
\bibitem [{\citenamefont {Bardyn}\ \emph {et~al.}(2018)\citenamefont {Bardyn}, \citenamefont {Wawer}, \citenamefont {Altland}, \citenamefont {Fleischhauer},\ and\ \citenamefont {Diehl}}]{PhysRevX.8.011035}%
  \BibitemOpen
  \bibfield  {author} {\bibinfo {author} {\bibfnamefont {C.-E.}\ \bibnamefont {Bardyn}}, \bibinfo {author} {\bibfnamefont {L.}~\bibnamefont {Wawer}}, \bibinfo {author} {\bibfnamefont {A.}~\bibnamefont {Altland}}, \bibinfo {author} {\bibfnamefont {M.}~\bibnamefont {Fleischhauer}},\ and\ \bibinfo {author} {\bibfnamefont {S.}~\bibnamefont {Diehl}},\ }\bibfield  {title} {\bibinfo {title} {Probing the topology of density matrices},\ }\href {https://doi.org/10.1103/PhysRevX.8.011035} {\bibfield  {journal} {\bibinfo  {journal} {Phys. Rev. X}\ }\textbf {\bibinfo {volume} {8}},\ \bibinfo {pages} {011035} (\bibinfo {year} {2018})}\BibitemShut {NoStop}%
\bibitem [{\citenamefont {Rainis}\ and\ \citenamefont {Loss}(2012)}]{PhysRevB.85.174533}%
  \BibitemOpen
  \bibfield  {author} {\bibinfo {author} {\bibfnamefont {D.}~\bibnamefont {Rainis}}\ and\ \bibinfo {author} {\bibfnamefont {D.}~\bibnamefont {Loss}},\ }\bibfield  {title} {\bibinfo {title} {Majorana qubit decoherence by quasiparticle poisoning},\ }\href {https://doi.org/10.1103/PhysRevB.85.174533} {\bibfield  {journal} {\bibinfo  {journal} {Phys. Rev. B}\ }\textbf {\bibinfo {volume} {85}},\ \bibinfo {pages} {174533} (\bibinfo {year} {2012})}\BibitemShut {NoStop}%
\bibitem [{\citenamefont {Coser}\ and\ \citenamefont {P{\'e}rez-Garc{\'\i}a}(2019)}]{coser2019classification}%
  \BibitemOpen
  \bibfield  {author} {\bibinfo {author} {\bibfnamefont {A.}~\bibnamefont {Coser}}\ and\ \bibinfo {author} {\bibfnamefont {D.}~\bibnamefont {P{\'e}rez-Garc{\'\i}a}},\ }\bibfield  {title} {\bibinfo {title} {Classification of phases for mixed states via fast dissipative evolution},\ }\href@noop {} {\bibfield  {journal} {\bibinfo  {journal} {Quantum}\ }\textbf {\bibinfo {volume} {3}},\ \bibinfo {pages} {174} (\bibinfo {year} {2019})}\BibitemShut {NoStop}%
\bibitem [{\citenamefont {Diehl}\ \emph {et~al.}(2011)\citenamefont {Diehl}, \citenamefont {Rico}, \citenamefont {Baranov},\ and\ \citenamefont {Zoller}}]{diehl2011topology}%
  \BibitemOpen
  \bibfield  {author} {\bibinfo {author} {\bibfnamefont {S.}~\bibnamefont {Diehl}}, \bibinfo {author} {\bibfnamefont {E.}~\bibnamefont {Rico}}, \bibinfo {author} {\bibfnamefont {M.~A.}\ \bibnamefont {Baranov}},\ and\ \bibinfo {author} {\bibfnamefont {P.}~\bibnamefont {Zoller}},\ }\bibfield  {title} {\bibinfo {title} {Topology by dissipation in atomic quantum wires},\ }\href@noop {} {\bibfield  {journal} {\bibinfo  {journal} {Nature physics}\ }\textbf {\bibinfo {volume} {7}},\ \bibinfo {pages} {971} (\bibinfo {year} {2011})}\BibitemShut {NoStop}%
\bibitem [{\citenamefont {Bardyn}\ \emph {et~al.}(2013)\citenamefont {Bardyn}, \citenamefont {Baranov}, \citenamefont {Kraus}, \citenamefont {Rico}, \citenamefont {Imamoglu}, \citenamefont {Zoller},\ and\ \citenamefont {Diehl}}]{Bardyn2013}%
  \BibitemOpen
  \bibfield  {author} {\bibinfo {author} {\bibfnamefont {C.-E.}\ \bibnamefont {Bardyn}}, \bibinfo {author} {\bibfnamefont {M.~A.}\ \bibnamefont {Baranov}}, \bibinfo {author} {\bibfnamefont {C.~V.}\ \bibnamefont {Kraus}}, \bibinfo {author} {\bibfnamefont {E.}~\bibnamefont {Rico}}, \bibinfo {author} {\bibfnamefont {A.}~\bibnamefont {Imamoglu}}, \bibinfo {author} {\bibfnamefont {P.}~\bibnamefont {Zoller}},\ and\ \bibinfo {author} {\bibfnamefont {S.}~\bibnamefont {Diehl}},\ }\bibfield  {title} {\bibinfo {title} {Topology by dissipation},\ }\href {https://doi.org/10.1088/1367-2630/15/8/085001} {\bibfield  {journal} {\bibinfo  {journal} {New J. Phys.}\ }\textbf {\bibinfo {volume} {15}},\ \bibinfo {pages} {085001} (\bibinfo {year} {2013})}\BibitemShut {NoStop}%
\bibitem [{\citenamefont {Deng}\ and\ \citenamefont {Yang}(2026)}]{deng2026dissipation}%
  \BibitemOpen
  \bibfield  {author} {\bibinfo {author} {\bibfnamefont {T.-S.}\ \bibnamefont {Deng}}\ and\ \bibinfo {author} {\bibfnamefont {F.}~\bibnamefont {Yang}},\ }\bibfield  {title} {\bibinfo {title} {Dissipation-driven topological phase transitions in open quantum systems independent of system hamiltonian},\ }\href {https://doi.org/10.1103/3wcr-sxtz} {\bibfield  {journal} {\bibinfo  {journal} {Physical Review B}\ }\textbf {\bibinfo {volume} {113}},\ \bibinfo {pages} {024312} (\bibinfo {year} {2026})}\BibitemShut {NoStop}%
\bibitem [{\citenamefont {Kraus}\ \emph {et~al.}(2008)\citenamefont {Kraus}, \citenamefont {B\"uchler}, \citenamefont {Diehl}, \citenamefont {Kantian}, \citenamefont {Micheli},\ and\ \citenamefont {Zoller}}]{PhysRevA.78.042307}%
  \BibitemOpen
  \bibfield  {author} {\bibinfo {author} {\bibfnamefont {B.}~\bibnamefont {Kraus}}, \bibinfo {author} {\bibfnamefont {H.~P.}\ \bibnamefont {B\"uchler}}, \bibinfo {author} {\bibfnamefont {S.}~\bibnamefont {Diehl}}, \bibinfo {author} {\bibfnamefont {A.}~\bibnamefont {Kantian}}, \bibinfo {author} {\bibfnamefont {A.}~\bibnamefont {Micheli}},\ and\ \bibinfo {author} {\bibfnamefont {P.}~\bibnamefont {Zoller}},\ }\bibfield  {title} {\bibinfo {title} {Preparation of entangled states by quantum markov processes},\ }\href {https://doi.org/10.1103/PhysRevA.78.042307} {\bibfield  {journal} {\bibinfo  {journal} {Phys. Rev. A}\ }\textbf {\bibinfo {volume} {78}},\ \bibinfo {pages} {042307} (\bibinfo {year} {2008})}\BibitemShut {NoStop}%
\bibitem [{\citenamefont {Mao}\ \emph {et~al.}(2024)\citenamefont {Mao}, \citenamefont {Yang},\ and\ \citenamefont {Zhai}}]{Mao2024}%
  \BibitemOpen
  \bibfield  {author} {\bibinfo {author} {\bibfnamefont {L.}~\bibnamefont {Mao}}, \bibinfo {author} {\bibfnamefont {F.}~\bibnamefont {Yang}},\ and\ \bibinfo {author} {\bibfnamefont {H.}~\bibnamefont {Zhai}},\ }\bibfield  {title} {\bibinfo {title} {Symmetry-preserving quadratic lindbladian and dissipation driven topological transitions in gaussian states},\ }\href {https://doi.org/10.1088/1361-6633/ad44d4} {\bibfield  {journal} {\bibinfo  {journal} {Rep. Prog. Phys.}\ }\textbf {\bibinfo {volume} {87}},\ \bibinfo {pages} {070501} (\bibinfo {year} {2024})}\BibitemShut {NoStop}%
\bibitem [{\citenamefont {Budich}\ \emph {et~al.}(2015)\citenamefont {Budich}, \citenamefont {Zoller},\ and\ \citenamefont {Diehl}}]{PhysRevA.91.042117}%
  \BibitemOpen
  \bibfield  {author} {\bibinfo {author} {\bibfnamefont {J.~C.}\ \bibnamefont {Budich}}, \bibinfo {author} {\bibfnamefont {P.}~\bibnamefont {Zoller}},\ and\ \bibinfo {author} {\bibfnamefont {S.}~\bibnamefont {Diehl}},\ }\bibfield  {title} {\bibinfo {title} {Dissipative preparation of chern insulators},\ }\href {https://doi.org/10.1103/PhysRevA.91.042117} {\bibfield  {journal} {\bibinfo  {journal} {Phys. Rev. A}\ }\textbf {\bibinfo {volume} {91}},\ \bibinfo {pages} {042117} (\bibinfo {year} {2015})}\BibitemShut {NoStop}%
\bibitem [{\citenamefont {Sieberer}\ \emph {et~al.}(2016)\citenamefont {Sieberer}, \citenamefont {Buchhold},\ and\ \citenamefont {Diehl}}]{Sieberer_2016}%
  \BibitemOpen
  \bibfield  {author} {\bibinfo {author} {\bibfnamefont {L.~M.}\ \bibnamefont {Sieberer}}, \bibinfo {author} {\bibfnamefont {M.}~\bibnamefont {Buchhold}},\ and\ \bibinfo {author} {\bibfnamefont {S.}~\bibnamefont {Diehl}},\ }\bibfield  {title} {\bibinfo {title} {Keldysh field theory for driven open quantum systems},\ }\href {https://doi.org/10.1088/0034-4885/79/9/096001} {\bibfield  {journal} {\bibinfo  {journal} {Reports on Progress in Physics}\ }\textbf {\bibinfo {volume} {79}},\ \bibinfo {pages} {096001} (\bibinfo {year} {2016})}\BibitemShut {NoStop}%
\bibitem [{\citenamefont {Su}\ \emph {et~al.}(2025)\citenamefont {Su}, \citenamefont {Sarma}, \citenamefont {Bintz}, \citenamefont {Kiely}, \citenamefont {Bao}, \citenamefont {Fisher},\ and\ \citenamefont {Xu}}]{pjs3-14cc}%
  \BibitemOpen
  \bibfield  {author} {\bibinfo {author} {\bibfnamefont {K.}~\bibnamefont {Su}}, \bibinfo {author} {\bibfnamefont {A.}~\bibnamefont {Sarma}}, \bibinfo {author} {\bibfnamefont {M.}~\bibnamefont {Bintz}}, \bibinfo {author} {\bibfnamefont {T.}~\bibnamefont {Kiely}}, \bibinfo {author} {\bibfnamefont {Y.}~\bibnamefont {Bao}}, \bibinfo {author} {\bibfnamefont {M.~P.~A.}\ \bibnamefont {Fisher}},\ and\ \bibinfo {author} {\bibfnamefont {C.}~\bibnamefont {Xu}},\ }\bibfield  {title} {\bibinfo {title} {Spin liquid and superconductivity emerging from steady states and measurements},\ }\href {https://doi.org/10.1103/pjs3-14cc} {\bibfield  {journal} {\bibinfo  {journal} {Phys. Rev. Lett.}\ }\textbf {\bibinfo {volume} {135}},\ \bibinfo {pages} {050403} (\bibinfo {year} {2025})}\BibitemShut {NoStop}%
\bibitem [{\citenamefont {Sarma}\ \emph {et~al.}(2025)\citenamefont {Sarma}, \citenamefont {Bao}, \citenamefont {Myerson-Jain}, \citenamefont {Kiely},\ and\ \citenamefont {Xu}}]{nn2m-w4vk}%
  \BibitemOpen
  \bibfield  {author} {\bibinfo {author} {\bibfnamefont {A.}~\bibnamefont {Sarma}}, \bibinfo {author} {\bibfnamefont {Y.}~\bibnamefont {Bao}}, \bibinfo {author} {\bibfnamefont {N.}~\bibnamefont {Myerson-Jain}}, \bibinfo {author} {\bibfnamefont {T.}~\bibnamefont {Kiely}},\ and\ \bibinfo {author} {\bibfnamefont {C.}~\bibnamefont {Xu}},\ }\bibfield  {title} {\bibinfo {title} {Effective conformal field theory generated from pure and dephased chern insulator},\ }\href {https://doi.org/10.1103/nn2m-w4vk} {\bibfield  {journal} {\bibinfo  {journal} {Phys. Rev. B}\ }\textbf {\bibinfo {volume} {112}},\ \bibinfo {pages} {085130} (\bibinfo {year} {2025})}\BibitemShut {NoStop}%
\bibitem [{\citenamefont {Budich}\ and\ \citenamefont {Diehl}(2015)}]{PhysRevB.91.165140}%
  \BibitemOpen
  \bibfield  {author} {\bibinfo {author} {\bibfnamefont {J.~C.}\ \bibnamefont {Budich}}\ and\ \bibinfo {author} {\bibfnamefont {S.}~\bibnamefont {Diehl}},\ }\bibfield  {title} {\bibinfo {title} {Topology of density matrices},\ }\href {https://doi.org/10.1103/PhysRevB.91.165140} {\bibfield  {journal} {\bibinfo  {journal} {Phys. Rev. B}\ }\textbf {\bibinfo {volume} {91}},\ \bibinfo {pages} {165140} (\bibinfo {year} {2015})}\BibitemShut {NoStop}%
\bibitem [{\citenamefont {You}\ \emph {et~al.}(2014)\citenamefont {You}, \citenamefont {Bi}, \citenamefont {Rasmussen}, \citenamefont {Slagle},\ and\ \citenamefont {Xu}}]{PhysRevLett.112.247202}%
  \BibitemOpen
  \bibfield  {author} {\bibinfo {author} {\bibfnamefont {Y.-Z.}\ \bibnamefont {You}}, \bibinfo {author} {\bibfnamefont {Z.}~\bibnamefont {Bi}}, \bibinfo {author} {\bibfnamefont {A.}~\bibnamefont {Rasmussen}}, \bibinfo {author} {\bibfnamefont {K.}~\bibnamefont {Slagle}},\ and\ \bibinfo {author} {\bibfnamefont {C.}~\bibnamefont {Xu}},\ }\bibfield  {title} {\bibinfo {title} {Wave function and strange correlator of short-range entangled states},\ }\href {https://doi.org/10.1103/PhysRevLett.112.247202} {\bibfield  {journal} {\bibinfo  {journal} {Phys. Rev. Lett.}\ }\textbf {\bibinfo {volume} {112}},\ \bibinfo {pages} {247202} (\bibinfo {year} {2014})}\BibitemShut {NoStop}%
\bibitem [{\citenamefont {Chaduteau}\ \emph {et~al.}(2026)\citenamefont {Chaduteau}, \citenamefont {Lee}, \citenamefont {Schindler},\ and\ \citenamefont {Prem}}]{chaduteau2026topologydecoherence}%
  \BibitemOpen
  \bibfield  {author} {\bibinfo {author} {\bibfnamefont {A.}~\bibnamefont {Chaduteau}}, \bibinfo {author} {\bibfnamefont {D.}~\bibnamefont {Lee}}, \bibinfo {author} {\bibfnamefont {F.}~\bibnamefont {Schindler}},\ and\ \bibinfo {author} {\bibfnamefont {A.}~\bibnamefont {Prem}},\ }\href {https://arxiv.org/abs/2607.07801} {\bibinfo {title} {Topology from decoherence}} (\bibinfo {year} {2026}),\ \Eprint {https://arxiv.org/abs/2607.07801} {arXiv:2607.07801 [quant-ph]} \BibitemShut {NoStop}%
\bibitem [{\citenamefont {Wang}\ \emph {et~al.}(2026{\natexlab{b}})\citenamefont {Wang}, \citenamefont {Kiely}, \citenamefont {Tell}, \citenamefont {Obermeyer}, \citenamefont {Barendregt}, \citenamefont {Bojović}, \citenamefont {Preiss}, \citenamefont {Sarma}, \citenamefont {Franz}, \citenamefont {Fisher}, \citenamefont {Xu},\ and\ \citenamefont {Bloch}}]{wang2026observationstrongtoweakspontaneoussymmetry}%
  \BibitemOpen
  \bibfield  {author} {\bibinfo {author} {\bibfnamefont {S.}~\bibnamefont {Wang}}, \bibinfo {author} {\bibfnamefont {T.~G.}\ \bibnamefont {Kiely}}, \bibinfo {author} {\bibfnamefont {D.}~\bibnamefont {Tell}}, \bibinfo {author} {\bibfnamefont {J.}~\bibnamefont {Obermeyer}}, \bibinfo {author} {\bibfnamefont {M.}~\bibnamefont {Barendregt}}, \bibinfo {author} {\bibfnamefont {P.}~\bibnamefont {Bojović}}, \bibinfo {author} {\bibfnamefont {P.~M.}\ \bibnamefont {Preiss}}, \bibinfo {author} {\bibfnamefont {A.}~\bibnamefont {Sarma}}, \bibinfo {author} {\bibfnamefont {T.}~\bibnamefont {Franz}}, \bibinfo {author} {\bibfnamefont {M.~P.~A.}\ \bibnamefont {Fisher}}, \bibinfo {author} {\bibfnamefont {C.}~\bibnamefont {Xu}},\ and\ \bibinfo {author} {\bibfnamefont {I.}~\bibnamefont {Bloch}},\ }\href {https://arxiv.org/abs/2604.16137} {\bibinfo {title} {Observation of strong-to-weak spontaneous symmetry breaking in a dephased fermi gas}} (\bibinfo {year} {2026}{\natexlab{b}}),\ \Eprint {https://arxiv.org/abs/2604.16137}
  {arXiv:2604.16137 [cond-mat.quant-gas]} \BibitemShut {NoStop}%
\bibitem [{\citenamefont {Kiely}\ and\ \citenamefont {Xu}(2025)}]{8vtp-khnq}%
  \BibitemOpen
  \bibfield  {author} {\bibinfo {author} {\bibfnamefont {T.~G.}\ \bibnamefont {Kiely}}\ and\ \bibinfo {author} {\bibfnamefont {C.}~\bibnamefont {Xu}},\ }\bibfield  {title} {\bibinfo {title} {Phase transition of topological index driven by dephasing},\ }\href {https://doi.org/10.1103/8vtp-khnq} {\bibfield  {journal} {\bibinfo  {journal} {Phys. Rev. B}\ }\textbf {\bibinfo {volume} {112}},\ \bibinfo {pages} {L201112} (\bibinfo {year} {2025})}\BibitemShut {NoStop}%
\bibitem [{\citenamefont {Fukui}\ \emph {et~al.}(2005)\citenamefont {Fukui}, \citenamefont {Hatsugai},\ and\ \citenamefont {Suzuki}}]{Fukui2005}%
  \BibitemOpen
  \bibfield  {author} {\bibinfo {author} {\bibfnamefont {T.}~\bibnamefont {Fukui}}, \bibinfo {author} {\bibfnamefont {Y.}~\bibnamefont {Hatsugai}},\ and\ \bibinfo {author} {\bibfnamefont {H.}~\bibnamefont {Suzuki}},\ }\bibfield  {title} {\bibinfo {title} {Chern numbers in discretized brillouin zone: Efficient method of computing (spin) hall conductances},\ }\href {https://doi.org/10.1143/JPSJ.74.1674} {\bibfield  {journal} {\bibinfo  {journal} {J. Phys. Soc. Jpn.}\ }\textbf {\bibinfo {volume} {74}},\ \bibinfo {pages} {1674} (\bibinfo {year} {2005})}\BibitemShut {NoStop}%
\bibitem [{\citenamefont {Fukui}\ and\ \citenamefont {Hatsugai}(2007)}]{PhysRevB.75.121403}%
  \BibitemOpen
  \bibfield  {author} {\bibinfo {author} {\bibfnamefont {T.}~\bibnamefont {Fukui}}\ and\ \bibinfo {author} {\bibfnamefont {Y.}~\bibnamefont {Hatsugai}},\ }\bibfield  {title} {\bibinfo {title} {Topological aspects of the quantum spin-hall effect in graphene: ${\mathrm{z}}_{2}$ topological order and spin chern number},\ }\href {https://doi.org/10.1103/PhysRevB.75.121403} {\bibfield  {journal} {\bibinfo  {journal} {Phys. Rev. B}\ }\textbf {\bibinfo {volume} {75}},\ \bibinfo {pages} {121403(R)} (\bibinfo {year} {2007})}\BibitemShut {NoStop}%
\bibitem [{\citenamefont {Gresch}\ \emph {et~al.}(2017)\citenamefont {Gresch}, \citenamefont {Autes}, \citenamefont {Yazyev}, \citenamefont {Troyer}, \citenamefont {Vanderbilt}, \citenamefont {Bernevig},\ and\ \citenamefont {Soluyanov}}]{gresch2017z2pack}%
  \BibitemOpen
  \bibfield  {author} {\bibinfo {author} {\bibfnamefont {D.}~\bibnamefont {Gresch}}, \bibinfo {author} {\bibfnamefont {G.}~\bibnamefont {Autes}}, \bibinfo {author} {\bibfnamefont {O.~V.}\ \bibnamefont {Yazyev}}, \bibinfo {author} {\bibfnamefont {M.}~\bibnamefont {Troyer}}, \bibinfo {author} {\bibfnamefont {D.}~\bibnamefont {Vanderbilt}}, \bibinfo {author} {\bibfnamefont {B.~A.}\ \bibnamefont {Bernevig}},\ and\ \bibinfo {author} {\bibfnamefont {A.~A.}\ \bibnamefont {Soluyanov}},\ }\bibfield  {title} {\bibinfo {title} {Z2pack: Numerical implementation of hybrid wannier centers for identifying topological materials},\ }\href@noop {} {\bibfield  {journal} {\bibinfo  {journal} {Physical Review B}\ }\textbf {\bibinfo {volume} {95}},\ \bibinfo {pages} {075146} (\bibinfo {year} {2017})}\BibitemShut {NoStop}%
\bibitem [{\citenamefont {Soluyanov}\ and\ \citenamefont {Vanderbilt}(2011)}]{PhysRevB.83.235401}%
  \BibitemOpen
  \bibfield  {author} {\bibinfo {author} {\bibfnamefont {A.~A.}\ \bibnamefont {Soluyanov}}\ and\ \bibinfo {author} {\bibfnamefont {D.}~\bibnamefont {Vanderbilt}},\ }\bibfield  {title} {\bibinfo {title} {Computing topological invariants without inversion symmetry},\ }\href {https://doi.org/10.1103/PhysRevB.83.235401} {\bibfield  {journal} {\bibinfo  {journal} {Phys. Rev. B}\ }\textbf {\bibinfo {volume} {83}},\ \bibinfo {pages} {235401} (\bibinfo {year} {2011})}\BibitemShut {NoStop}%
\bibitem [{\citenamefont {Alba}\ \emph {et~al.}(2011)\citenamefont {Alba}, \citenamefont {Fernandez-Gonzalvo}, \citenamefont {Mur-Petit}, \citenamefont {Pachos},\ and\ \citenamefont {Garc{\'\i}a-Ripoll}}]{alba2011seeing}%
  \BibitemOpen
  \bibfield  {author} {\bibinfo {author} {\bibfnamefont {E.}~\bibnamefont {Alba}}, \bibinfo {author} {\bibfnamefont {X.}~\bibnamefont {Fernandez-Gonzalvo}}, \bibinfo {author} {\bibfnamefont {J.}~\bibnamefont {Mur-Petit}}, \bibinfo {author} {\bibfnamefont {J.~K.}\ \bibnamefont {Pachos}},\ and\ \bibinfo {author} {\bibfnamefont {J.~J.}\ \bibnamefont {Garc{\'\i}a-Ripoll}},\ }\bibfield  {title} {\bibinfo {title} {Seeing topological order in time-of-flight measurements},\ }\href@noop {} {\bibfield  {journal} {\bibinfo  {journal} {Physical review letters}\ }\textbf {\bibinfo {volume} {107}},\ \bibinfo {pages} {235301} (\bibinfo {year} {2011})}\BibitemShut {NoStop}%
\bibitem [{\citenamefont {Rubio-Garc{\'\i}a}\ \emph {et~al.}(2020)\citenamefont {Rubio-Garc{\'\i}a}, \citenamefont {Self}, \citenamefont {Garc{\'\i}a-Ripoll},\ and\ \citenamefont {Pachos}}]{rubio2020seeing}%
  \BibitemOpen
  \bibfield  {author} {\bibinfo {author} {\bibfnamefont {A.}~\bibnamefont {Rubio-Garc{\'\i}a}}, \bibinfo {author} {\bibfnamefont {C.~N.}\ \bibnamefont {Self}}, \bibinfo {author} {\bibfnamefont {J.~J.}\ \bibnamefont {Garc{\'\i}a-Ripoll}},\ and\ \bibinfo {author} {\bibfnamefont {J.~K.}\ \bibnamefont {Pachos}},\ }\bibfield  {title} {\bibinfo {title} {Seeing topological edge and bulk currents in time-of-flight images},\ }\href@noop {} {\bibfield  {journal} {\bibinfo  {journal} {Physical Review B}\ }\textbf {\bibinfo {volume} {102}},\ \bibinfo {pages} {041123} (\bibinfo {year} {2020})}\BibitemShut {NoStop}%
\bibitem [{\citenamefont {Datta}(1995)}]{Datta_1995}%
  \BibitemOpen
  \bibfield  {author} {\bibinfo {author} {\bibfnamefont {S.}~\bibnamefont {Datta}},\ }\href {https://doi.org/10.1017/CBO9780511805776} {\emph {\bibinfo {title} {Electronic Transport in Mesoscopic Systems}}},\ Cambridge Studies in Semiconductor Physics and Microelectronic Engineering\ (\bibinfo  {publisher} {Cambridge University Press},\ \bibinfo {address} {Cambridge},\ \bibinfo {year} {1995})\BibitemShut {NoStop}%
\bibitem [{\citenamefont {Groth}\ \emph {et~al.}(2014)\citenamefont {Groth}, \citenamefont {Wimmer}, \citenamefont {Akhmerov},\ and\ \citenamefont {Waintal}}]{Groth_2014}%
  \BibitemOpen
  \bibfield  {author} {\bibinfo {author} {\bibfnamefont {C.~W.}\ \bibnamefont {Groth}}, \bibinfo {author} {\bibfnamefont {M.}~\bibnamefont {Wimmer}}, \bibinfo {author} {\bibfnamefont {A.~R.}\ \bibnamefont {Akhmerov}},\ and\ \bibinfo {author} {\bibfnamefont {X.}~\bibnamefont {Waintal}},\ }\bibfield  {title} {\bibinfo {title} {Kwant: a software package for quantum transport},\ }\href {https://doi.org/10.1088/1367-2630/16/6/063065} {\bibfield  {journal} {\bibinfo  {journal} {New Journal of Physics}\ }\textbf {\bibinfo {volume} {16}},\ \bibinfo {pages} {063065} (\bibinfo {year} {2014})}\BibitemShut {NoStop}%
\end{thebibliography}%

\end{document}